\documentclass[superscriptaddress,nofootinbib,pre,twocolumn]{revtex4}

\usepackage{amsmath,amssymb}
\usepackage{graphicx}
\usepackage{bm}
\usepackage[colorlinks,linkcolor=blue,urlcolor=cyan,citecolor=blue,
anchorcolor=blue]{hyperref}
\usepackage[T1]{fontenc}
\usepackage{mathrsfs}
\usepackage{color}

\begin{document}

\title{Behavior of passive polymeric tracers of different topologies in
a dilute bath of active Brownian particles}

\author{Ramanand Singh Yadav}
\affiliation{Department of Chemistry, Indian Institute of Technology Bombay,
Mumbai 400076, India}
\author{Ralf Metzler}
\email{rmetzler@uni-potsdam.de (Corresponding author)}
\affiliation{Institute of Physics and Astronomy, University of Potsdam,
14476 Potsdam, Germany}
\affiliation{Asia Pacific Centre for Theoretical Physics, Pohang 37673,
Republic of Korea}
\author{Rajarshi Chakrabarti}
\email{rajarshi@chem.iitb.ac.in (Corresponding author)}
\affiliation{Department of Chemistry, Indian Institute of Technology Bombay,
Mumbai 400076, India}

\begin{abstract}
Using computer simulations in two dimensions we investigate the dynamics
and structure of passive polymeric tracer with different topologies immersed
in a low-density active particle bath. One of the key observations is that
polymer exhibit faster dynamics compared to passive colloidal particles at
high activity, for the same particle density, in both linear and star polymer
topologies. This enhanced motion is attributed to the accumulation of active
particles, which induces prolonged and persistent movement of the polymer.
Further analysis reveals that star polymers exhibit more complex and intriguing
behavior than their linear counterparts. Notably, the accumulation of active
particles promotes the pairing of arms in star polymers. For instance, a
three-armed star polymer adopts a conformation similar to a linear polymer
with two-arms due to this pairing as a result, at high activity, the dynamics
of both the polymers converge.  Finally, we explore the dynamics of a linear
polymer with the same total number of beads as the star polymer. Interestingly,
at high activity---where arm pairing in the star polymer is significant---the
star polymer demonstrates faster dynamics than the linear polymer, despite
having the identical number of beads. These findings contribute to a broader
understanding of the interactions between active and passive components
of varying topologies in dilute systems and highlight their potential for
innovative applications ranging from materials science to biomedicine.

\end{abstract}

\maketitle

\section{Introduction}
\label{sec:Introduction}
\label{sec:Intro3}

Active matter is constituted of small units that convert their internal
energy---typically replenished periodically from their environment---into
directed motion. Unlike passive systems, active matter exhibits non-thermal
fluctuations, breaks detailed balance, and violates the fluctuation-dissipation
relation, keeping the system inherently out of equilibrium
\cite{ramaswamy2010mechanics,bechinger2016active}. These characteristics
enable intriguing behaviors and potential applications, such as targeted
cargo transport and efficient delivery in both biological (in vivo) and
laboratory (in vitro) environments. Several studies have been dedicated to
understanding how active forces interact with thermodynamic forces
\cite{wang2013small,mallory2018active,klapp2016collective,ramaswamy2010mechanics,marchetti2016minimal,menzel2015tuned,bialke2015active,zottl2016emergent},
and to exploring ways to harness these forces for specific microscale tasks
\cite{di2016controlled,frangipane2018dynamic}.

The coexistence of active and passive elements is a key feature of many living
systems. For instance, the motility of microorganisms in active systems plays a
crucial role in nutrient mixing and maintaining ecological balance in aquatic
environments \cite{kurtuldu2011enhancement,stocker2012ecology}.
The enhanced diffusion of passive components in an active bath affects various
passive entities such as enzymes, granules, and extracellular products
\cite{caspi2000enhanced}.
These components are influenced by cyclic conformational changes in biomolecules
such as DNA and RNA during active transcription processes, which are fueled by
ATP \cite{guo2014probing}.

Recent studies have explored the statistical mechanics of active agents in a
passive bath \cite{majumdar2022exactly,martin2021statistical,
theeyancheri2023active} as well as passive agents in an active bath
\cite{pietzonka2017entropy,chaki2018entropy,chaki2019effects,chaki2019enhanced,
ye2020active,shea2022passive,goswami2023trapped,grossmann2024non,
anand2020conformation,shin2015facilitation,kaiser2015does,chaki2019enhanced,
goswami2022reconfiguration,samanta2016chain,aporvari2020anisotropic,
ghosh2014dynamics,osmanovic2017dynamics}. In contrast to spherical passive
colloidal particles in an active bath, asymmetric passive agents---such as
V-shaped wedges---exhibit a more pronounced directed motion. This occurs
because active particles tend to accumulate in the cusp of the wedge, exerting
an asymmetric pressure on its walls and driving a persistent movement
\cite{kaiser2014transport,angelani2010geometrically}. In a passive
bath, in contrast, the pressure exerted by passive particles is isotropic,
resulting in no net directional motion.

The simple model system of a passive tracer in an active bath has been explored
both numerically and experimentally by various groups \cite{smallenburg2015swim,
angelani2010geometrically,kaiser2014transport}. It has been observed that maximal
efficiency occurs at an optimal number of active particles. At high packing
fractions, the bath becomes jammed, which restricts the mobility of the carrier
object \cite{yadav2024passive}. Conversely, at low packing fractions, there is
insufficient accumulation of active particles near the object to induce
significant motion. Unlike hard asymmetric objects, soft and flexible objects
in an active bath display intriguing behaviors. Several studies have
investigated the interactions between active particles and deformable passive
objects, as well as the behavior of active polymers \cite{winkler2017active,
mallory2014curvature}. A linear polymer immersed in a bath of active Brownian
particles (ABPs) adopts different conformations depending on its stiffness. For
instance, a semiflexible linear polymer forms transient hairpin structures due
to the accumulation of ABPs in regions of high curvature, which promotes
polymer shrinkage. In contrast, a fully flexible polymer tends to expand with
increasing activity \cite{chaki2019enhanced,shin2015facilitation,
goswami2022reconfiguration}. Similar to hard passive asymmetric objects,
flexible linear polymers also exhibit enhanced dynamics, driven by the
localized accumulation of active particles in highly curved ("parachute-shaped")
regions \cite{shin2015facilitation}. While previous studies have focused on
linear polymer topologies, less attention has been given to branched and
star-shaped polymers in active baths. Star-shaped polymers, in which multiple
arms are connected at a central core, have biological counterparts in the form
of certain viruses, such as astroviruses \cite{perot2017astrovirus}. Some
studies have also engineered star-shaped molecules to mimic viral properties,
investigating their dynamics \cite{de2017rod,zavala2023m13}. Moreover, the
tunable properties of star-shaped polymers have made them an attractive candidate
for drug delivery applications \cite{yang2017nano,sulistio2011folic,
liu2012synthesis,biffi2013phase,liu2004dna}. While computational studies have
examined the behavior of star polymers in passive baths (both good and bad
solvents) \cite{taddese2015thermodynamics}, investigating their conformational
changes and dynamics, there is still limited research on the behavior of
star-shaped polymers in an active bath \cite{yadav2024passive}.

In this work, we explore the dynamics and conformational changes of star
polymers and compare their behavior to that of tracers with other topologies
in a dilute bath of ABPs using computer simulations in two dimensions. We
observe an intriguing conformational behavior of the star-shaped polymers in
such an environment. Notably, beyond a threshold value of the active force,
a "folded" conformation emerges for the star polymer with three arms, such that
two arms are paired while the third remains at a distance. This occurs due
to the asymmetric accumulation of active particles. In previous studies,
a similar behavior has also been observed in systems of flexible polymers,
where increased activity at low packing fractions promotes the pairing of
two polymer chains \cite{gandikota2022effective}. In contrast, at relatively
high densities, active particles can form dynamic crystalline structures,
establishing a stable, oscillating bridge between particles. This bridge gives
rise to significant long-range dynamic repulsion during wetting near confining
boundaries. However, as the density decreases, the bridge gradually breaks
down, resulting in an intriguing transition to long-range dynamic depletion
attraction \cite{yadav2024passive,ni2015tunable,angelani2011effective}. The
pairing of the arms, together with the accumulation of active particles,
drives prolonged directed motion. This is evident from the increased mean
squared displacement (MSD), the positive temporal velocity autocorrelation
function (VACF), and the enhanced persistence time with increasing activity.

The described arm pairing in star polymers is more favorable and occurs at
lower activity levels for semiflexible polymers as compared to flexible ones.
When comparing the behavior of a star polymer with that of passive colloidal
particles---whose size is similar to that of a single bead of the polymer---we
find that the dynamics of the star polymer's center of mass is faster and
more persistent at high activity. This is attributed to the accumulation
of active particles and the pairing of arms. The difference is evident in
the higher MSD of the star polymer's center of mass compared to that of the
colloidal particle at long times and high activity. Additionally, the star
polymer exhibits a larger scaling exponent of the MSD and a more pronounced
positive temporal VACF under these conditions. Interestingly, at high activity
the conformation of the star polymer---in which two arms are paired and the
third is extended---resembles that of a linear polymer. To gain a clearer
understanding, we compare the behavior of a three-armed star polymer (with 61
beads) with that of a linear polymer (41 beads), and our findings suggest that
their dynamics converge at high activity. Additionally, we compare the star
polymer with a linear polymer containing the same number of beads. The results
indicate that, at high activity---where arm pairing in the star polymer is
significant---the dynamics of the star polymer are slightly faster and more
persistent than those of the linear polymer, despite having the same number
of beads. This difference can be explained by the distribution of the radius
of gyration, as the linear polymer exhibits a significantly larger radius of
gyration compared to the star polymer. Additionally, the time evolution of the
VACF shows that the star polymer maintains a higher correlation as compared
to the linear polymer.

This paper is organized as follows. In Section \ref{sec:Model3}, we present
the model and simulation details. Results and discussion are presented
in Section \ref{sec:Result3} and lastly, we conclude the paper in Section
\ref{sec:conclusion}. Additional information is compiled in the Appendix.

\section{Model and simulation details}
\label{sec:Simulation}
\label{sec:Model3} 

\begin{figure*}
\includegraphics[width=0.98\linewidth]{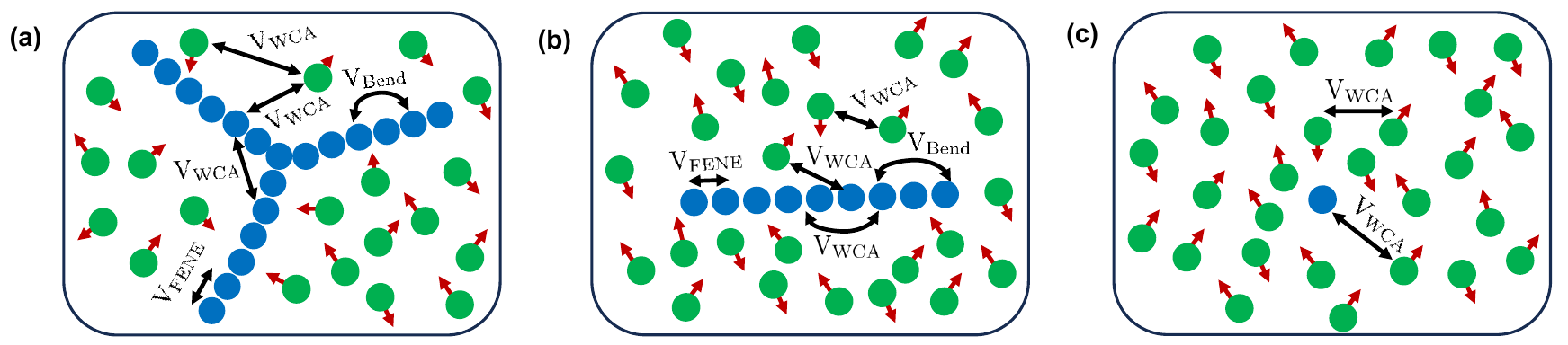}
\caption{Schematic of the model system (not to scale): (a) a single star
polymer with three arms (blue). (b) A passive linear polymer (blue). (c)
A single passive particle (blue). Each of these systems is simulated
separately while immersed in a dilute bath of ABPs (green). The red arrows
indicate the instantaneous directions of the ABPs. Pairwise non-bonded
interactions among the polymer beads, between the polymer beads and ABPs,
between passive particles and ABPs, and among the ABPs themselves are modeled
using the Weeks-Chandler-Andersen (WCA) potential, represented by double-headed
arrows. Additionally, the neighboring beads of the polymer are connected via
the finitely extensible nonlinear elastic (FENE) potential, and a bending
potential is applied to restrict polymer bending.}
\label{fig:image1.png}
\end{figure*}

We carry out coarse-grained computer simulations for a passive tracer embedded
in a bath of ABPs in two dimensions with periodic boundary conditions in all
directions. The considered tracers include passive colloidal particles, linear
polymers with 41 and 61 monomers, and a star polymer with three arms, in which
each arm consists of 20 monomers and is connected with a central monomer. The
diameter of the individual tracer particles and the monomers of the polymeric
objects is $\sigma$, the same size as the bath ABPs. To set up the system, we
fix the passive tracer in a square box of length $120\sigma$ populated with
ABPs. A schematic of our system is shown in Fig.~\ref{fig:image1.png}. In the
simulations we fixed the packing fraction of the bath to $\phi=0.06$, where we
defined $\phi=N_{\mathrm{ABP}}A_{\mathrm{ABP}}/[120\sigma\times120\sigma]$.
Here $N_{\mathrm{ABP}}=1101$ is the number of ABPs and $A_{\mathrm{ABP}}=\pi
(\frac{\sigma}{2})^2$ is the area of a single ABP.

The dynamics of the position $\mathbf{r}_i(t)$ of the $i$th particle of mass
$m$ is described by the Langevin equation
\begin{equation}
m\frac{d^2\mathbf{r}_i(t)}{dt^2}=-\gamma\frac{d\mathbf{r}_i}{dt}-\sum_j\nabla
V(\mathbf{r}_i-\mathbf{r}_j)+\mathbf{f}_i(t)+F\mathbf{e}(\boldsymbol{\theta}_i).
\label{eq:langevineq}
\end{equation}
In the overdamped limit, the inertial term $m\frac{d^2\mathbf{r}_i(t)}{dt^2}$
is negligibly small as compared to the drag force $\gamma\frac{d \mathbf{r}_i}{
dt}$, where $\gamma$ is the friction coefficient. To ensure that the system is
practically overdamped, we took the comparatively high value of $\gamma=200$.
The $k$\textsuperscript{th} component $f_{ik}(t)$ of the thermal force $\mathbf{f}_i$ on particle
$i$ is taken as an independent Gaussian white noise with zero mean and variance
$\langle f_{ik}(t)f_{jl}(t')\rangle=2\gamma k_BT\delta_{ij}\delta{kl}\delta(
t-t')$, where $\delta_{ij}$ is the Kronecker delta and $\delta(t)$ is the Dirac
$\delta$-function. Moreover, $k_BT$ represents the thermal energy in terms of
the Boltzmann constant $k_B$ and the absolute temperature $T$. The ABPs are
simulated as disks of diameter $\sigma$ moving under the action of a constant
force $F$ along a predefined orientation vector
\begin{equation}
\mathbf{e}(\boldsymbol{\theta}_i)=\{\cos(\theta_i),\sin(\theta_i)\}.
\end{equation}
$F\mathbf{e}$ is thus the active force ("propulsion strength") of magnitude $F$
of the ABPs that drives the system out of equilibrium. The dimensionless
P{\'e}clet number can be used to express this propulsion strength relative to
the thermal particle speed as:
\begin{equation}
\mathrm{Pe}=\frac{v\sigma}{D}=\frac{F\sigma}{k_BT} 
\label{eq:peclet}
\end{equation}
such that $\mathrm{Pe}=0$ for passive tracers and $\mathrm{Pe}>0$
for the ABPs.

The orientation $\boldsymbol{\theta}_i$ of the velocity of the $i$\textsuperscript{th} ABP is
a function of time according to the stochastic equation
\begin{equation}
\frac{d\boldsymbol{\theta}_i}{dt}=\sqrt{2D_R}\times\pmb{\eta}(t).
\label{eq:example}
\end{equation}
Here, $D_R$ is the rotational diffusion coefficient, which is related to the
persistence time $\tau_R$ of an ABP by the relation $\tau_R=1/D_R$
\cite{howse2007self}.
The term $\pmb{\eta}(t)$ represents a zero-mean Gaussian random vector with
unit variance. We again assume component-wise and particle-particle independence,
i.e.,
\begin{equation}
\langle\eta_{ik}(t)\eta_{jl}(t')\rangle=2D_R\delta_{kl}\delta_{ij}\delta(t-t'),
\end{equation}
where $\eta_{ik}(t)$ is the $k$\textsuperscript{th} component of the noise acting on particle $i$.

The total pairwise interaction potential $V(r_i-r_j)$ is defined as $V=V_{
\mathrm{WCA}}+V_{\mathrm{FENE}}+V_{\mathrm{Bend}}$. Here, the monomers of the
polymers are connected by the FENE potential
\begin{equation}
V_{\mathrm{FENE}}(r_{ij})=\left\{\begin{array}{ll}-\frac{Kr_{\mathrm{max}}
^2}{2}\log\left(1-\left[{\frac{r_{ij}}{r_{\mathrm{max}}}}\right]^2\right),
&r_{ij}\leq r_{\mathrm{max}}\\
=\infty,&\mbox{otherwise}\end{array}\right.
\label{eq:FENE}
\end{equation}
where $r_{ij}=\vert\mathbf{r}_i-\mathbf{r}_j\vert$ represents the separation
between monomers $i$ and $j$ (with position vectors $\mathbf{r}_i$ and
$\mathbf{r}_j$, respectively) of the polymer; $r_{\mathrm{max}}=1.5\sigma$ is
taken as the upper limit of $r_{ij}$ and $K$ denotes the force constant, which
accounts for the stiffness of the polymer. To restrict the bending of the
polymer we introduce the bending potential 
\begin{equation}
V_{\mathrm{Bend}}=\kappa(1-\cos(\phi_{\mathrm{bend}})),
\label{eq:bend}
\end{equation}
where $\kappa=75$ (in units of $k_BT$) for the bending coefficient and $\phi
_{\mathrm{bend}}$ is the angle between two successive bond vectors of the
polymer. We employ purely repulsive pairwise non-bonded interactions between
the monomers of the polymers, modeled in terms of the Weeks-Chandler-Andersen
(WCA) potential \cite{weeks1971role}
\begin{widetext}
\begin{equation}
V_{\mathrm{WCA}}(r_{ij})=\left\{\begin{array}{ll}4\epsilon_{ij}\left(\left[
\frac{\sigma_{ij}}{r_{ij}}\right]^{12}-\left[\frac{\sigma_{ij}}{r_{ij}}
\right]^6\right)+\epsilon_{ij},&\mbox{if }r_{ij}\leq2^{1/6}\sigma_{ij}\\
=0,&\mbox{otherwise}\end{array}\right..
\label{eq:WCA_3}
\end{equation}
\end{widetext}
Here $\epsilon_{ij}$ is the strength of the interaction with the effective
interaction diameter $\sigma_{ij}=(\sigma_i+\sigma_j)/2$. 

All simulations are performed using a Langevin thermostat, and the equation of
motion is integrated using the velocity Verlet algorithm in each time step
\cite{verlet1967computer}.
We initialize the system by placing the passive tracer in the center of the
bath and relaxing the initial configuration for $10^7$ steps. The simulations
are carried out for $6\times10^8$ steps, where the integration time step is
taken to be $5\times10^{-4}$, and the positions of the monomers are recorded
every $100$th step. The simulations are carried out using LAMMPS
\cite{plimpton1995fast}, a freely available open-source molecular dynamics
package. In our simulations, we measure the length in units of $\sigma$ and
the energy in units of thermal energy $\epsilon=k_BT$. All other physical
quantities are therefore presented in dimensionless units expressed in terms
of these fundamental units $\sigma$, $\epsilon$, and $m$. Details of the model
parameters are given in Tab.~\ref{tab:table4}.

\begin{table}
\caption{\label{tab:table4} Model Parameters used in our simulations.}
\begin{ruledtabular}
\begin{tabular}{cc}
Parameter & Value \\
\hline
$\sigma_{\mathrm{ABP}}$ & $\sigma$  \\
$\sigma_{\mathrm{beads}}$ & $\sigma$  \\
$N_{\mathrm{ABP}}$ & 1100 $(\phi = 0.06)$\\
$\frac{m}{\gamma}$ & $5\times10^{-3}$  \\
$k_BT$& 1 \\
$\Delta t$ & $5\times 10^{-4}$ \\
$\mathrm{Pe}$ & 0, 5, 10, 15, 20,\\
& 30, 40, 50, 60\\
\end{tabular}
\end{ruledtabular}
\end{table}

\section{Results and discussion}
\label{sec:Results}
\label{sec:Result3}

\subsection{Behavior of a three-armed star polymer in a dilute active bath}

Soft particles represent a class of materials with a "dual nature", in the
sense that they exhibit properties that fall in between those of hard colloids
and polymer coils \cite{vlassopoulos2014tunable}. Typical examples include
microgels \cite{wei2013mechanism}, micelles \cite{laurati2005starlike}, and
star polymers \cite{laurati2005starlike,gupta2015dynamic}, whose phase
behavior and dynamics are strongly influenced by their degree of softness. In
this context, multiarm star-shaped systems---with adjustable arm number, length,
and flexibility---offer a versatile platform for investigating how varying
degrees of "softness" influence the dynamic behavior, due to their tunable
deformability and potential for interdigitation. In literature it was studied
how the softness of a material can be modulated by the doping with active
particles \cite{zhou2024active}, and can influence the conformation and
dynamics of the polymer, which has been extensively studied for linear polymers
\cite{winkler2017active,shin2015facilitation,mallory2014curvature}. Here, in
this work, we explore the behavior of a three-armed star polymer in a dilute
bath of active particles, specifically, the dynamics, and the conformational
changes and observe some intriguing behavior.

\subsubsection{Conformational changes of the star polymer}

In contrast to linear polymers, star polymers exhibit a more intricate
architecture due to their multiple arms \cite{laurati2005starlike,
gupta2015dynamic,ding2016flow,nagarajan2019flow,erwin2010dynamics}.
To characterize their structural features, we compute the pairwise
center-of-mass separation
\begin{equation}
\mathbf{R}_{A_i\mathrm{-}A_j}=\sqrt{(x_{\mathrm{com}(i)}-x_{\mathrm{com}(j)})^2
+(y_{\mathrm{com}(i)}-y_{\mathrm{com}(j)})^2}.
\end{equation}
for $i,j=1,2,3$ between the three arms of a three-armed flexible ($\kappa=0$)
star polymer.

\begin{figure*}
\centering
\includegraphics[width=0.9\linewidth]{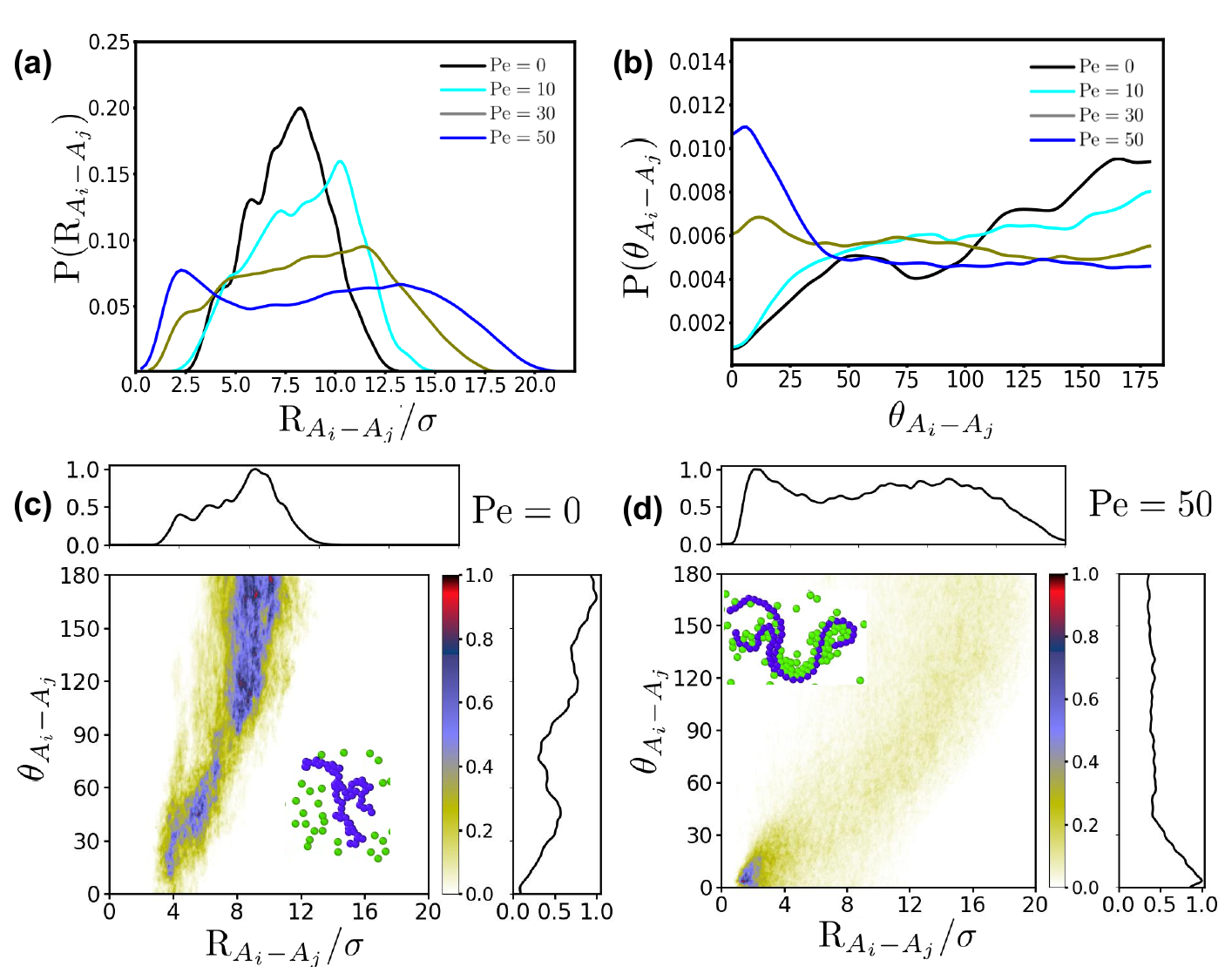}
\caption{Probability density functions of (a) the arm-arm separation
$\mathrm{R}_{A_i\mathrm{-}A_j}$ and (b) the arm-arm angle $\theta_{A_i\mathrm{
-}A_j}$ for different values of the P{\'e}clet number $\mathrm{Pe}$. Panels
(c) and (d) show the 1D and 2D probability of $\mathrm{R}_{A_i\mathrm{-}A_j}$
and $\theta_{A_i\mathrm{-}A_j}$ for $\mathrm{Pe}=0$ and $\mathrm{Pe}=50$,
respectively. Snapshots of conformations of the star polymer and associated
ABPs shown in panels (c) and (d) demonstrate the accumulation of ABPs induced
by higher activity.}
\label{fig:image2.png}
\end{figure*}

Fig.~\ref{fig:image2.png}(a) presents the probability density of this
separation $\mathrm{R}_{A_i\mathrm{-}A_j}$ for all possible combinations
of $i$ and $j$ ($i,j=1,2,3$), for varying P{\'e}clet number $\mathrm{Pe}$ of
the surrounding active bath particles. As $\mathrm{Pe}$ increases (we vary
it in the range $\mathrm{Pe}=0\ldots30)$, the distribution initially broadens,
indicating enhanced fluctuations in the arm separations as sown in
Movie\_S1 and Movie\_S2 (see App.~\ref{appf}).
At high activity ($\mathrm{Pe}=50$), a pronounced peak appears at smaller
values of $\mathrm{R}_{A_i\mathrm{-}A_j}$, followed by a broad distribution of
the population at larger separations. The peak at lower $\mathrm{R}_{A_i
\mathrm{-}A_j}$ does not occur exactly at $\mathrm{R}_{A_i\mathrm{-}A_j}/
\sigma=1.0$, due to the self-avoidance of the polymer beads and the
accumulation of some ABPs between the arms \cite{gandikota2022effective}. The
bimodality of the distribution is characteristic for a transient pairing of
two arms, while the third remains spatially separated, fluctuating over time
as shown in Movie\_S3 (see App.~\ref{appf}). To gain further insight into this
pairing behavior, we calculate the angle
$\theta_{A_i\mathrm{-}A_j}$,
\begin{equation}
\theta_{A_i\mathrm{-}A_j}=\cos^{-1}\left(\frac{\mathbf{A}_i\cdot\mathbf{A}_j}{
|\mathbf{A}_i||\mathbf{A}_j|}\right)
\end{equation}
between the arm vectors, where $\mathbf{A}_i=(x_i^{\mathrm{term}}-x_i^{
\mathrm{c}},y_i^{\mathrm{term}}-y_i^{\mathrm{c}})$, and $\mathbf{A}_j$ is
defined analogously. Here, the superscript "c" refers to the central bead
of the star polymer, and "term" denotes the terminal beads of the polymer.

As shown in Fig.~\ref{fig:image2.png}(b), the distribution of the arm-arm
angle $\theta_{A_i\mathrm{-}A_j}$ develops a sharp peak at lower angles with
increasing $\mathrm{Pe}$, particularly between $\mathrm{Pe}=30$ and $\mathrm{Pe}
=50$. This indicates the emergence of a preferential alignment or pairing
between two arms. A broader component at larger angles persists, reflecting
the variability in the orientation of the third arm.

Fig.~\ref{fig:image2.png}(c) and (d) display the joint probability density of
$\mathrm{R}_{A_i\mathrm{-}A_j}$ and $\theta_{A_i\mathrm{-}A_j}$ at $\mathrm{Pe}
=0$ and $\mathrm{Pe}=50$, respectively. At equilibrium ($\mathrm{Pe}=0$), the
distribution is centered around moderate values of $(\mathrm{R}_{A_i\mathrm{-}
A_j})$ with a broad angular spread, consistent with a disordered, coil-like
conformation. In contrast, under strong activity conditions ($\mathrm{Pe}=50$),
the distribution becomes varied, with a significant population concentrated at
low values of both $\mathrm{R}_{A_i\mathrm{-}A_j}$ and $\theta_{A_i\mathrm{-}
A_j}$, indicative of a paired-arm configuration. A secondary population at
higher values of both parameters corresponds to the fluctuating, unpaired
third arm. These results suggest that at high activity, the star polymer indeed
adopts a dynamic-induced asymmetric conformation wherein two arms transiently
associate, effectively forming a linear structure with the third arm extending
outward.

\begin{figure*}
\includegraphics[width=0.70\linewidth]{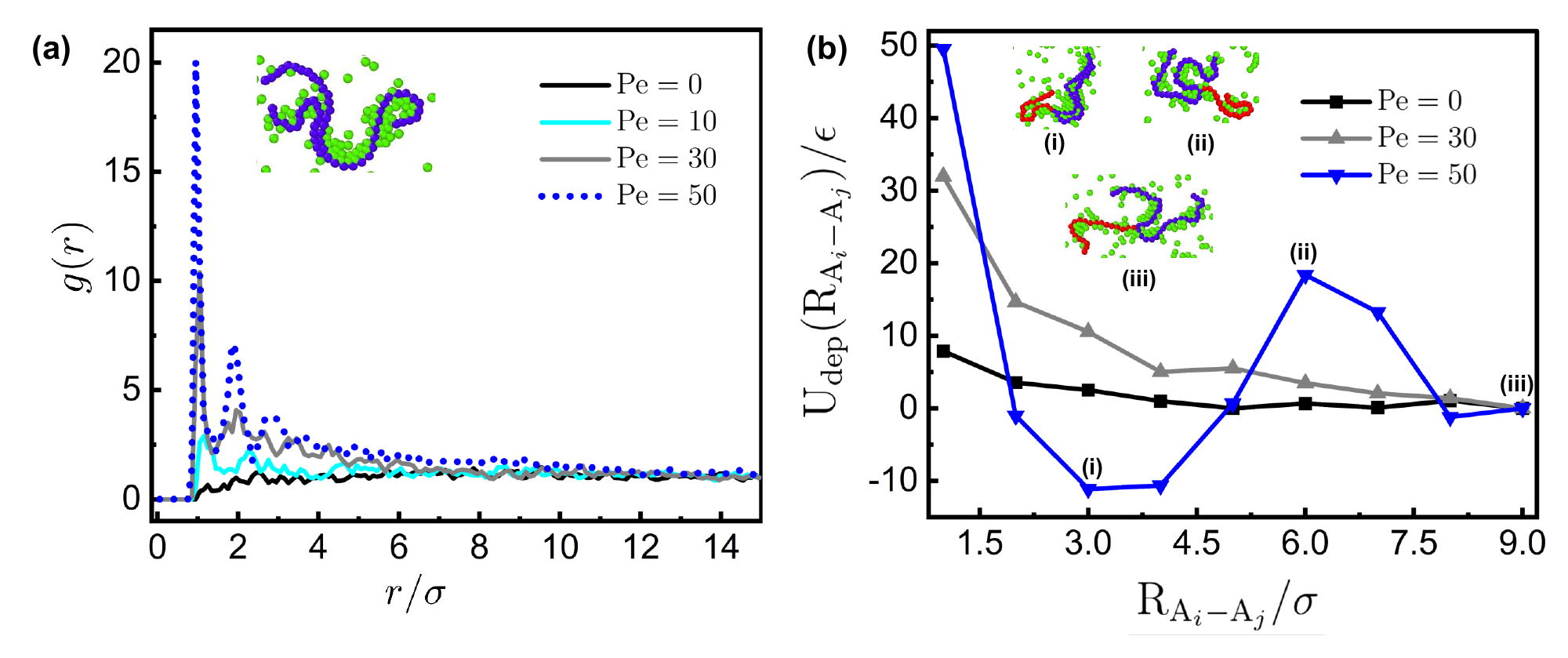}
\caption{Radial distribution function $g(r)$ of ABPs
surrounding (a) a star polymer for different $\mathrm{Pe}$. The inset in (a)
shows a snapshot of the system with the star polymer immersed in an active
bath at $\mathrm{Pe}=50$. (b) Effective interaction potential between a pair of
arms (blue) of the star polymer as a function of their separation $\mathrm{
R}_{A_i\mathrm{-}A_j}$ for different values of $\mathrm{Pe}$. The insets in (b)
show the conformations of the star polymer for a given separation $\mathrm{R}
_{A_i\mathrm{-}A_j}$: (i) $3.0$, (ii) $6.0$, and (iii) $9.0$.}
\label{fig:image3.png}
\end{figure*}

The accumulation of active particles near a polymer has a significant effect on
the conformation it adopts in the active bath. For instance, conformations such
as the bent structure of a semiflexible linear polymer or the looping of an
end-attractive linear polymer are strongly influenced by the presence of active
particles \cite{shin2015facilitation,harder2014activity,yadav2024passive}. To
understand the behavior of a star polymer, with its much complex structure as
compared to its linear counterpart, in an active bath, we calculate the radial
distribution function $g(r)$, of ABPs around the polymer,
as shown in Fig.~\ref{fig:image3.png}(a). We observe a noticeable accumulation
of ABPs in the vicinity of the chain. This accumulation becomes more pronounced
with increasing $\mathrm{Pe}$, reflected in the increasing height of the peak
of $g(r)$ at $r=1$. We also observe that with
bath increasing activity, not only does the peak at $r=1$ become more
pronounced, but also a secondary peak emerges at a larger distance. The inset
in Fig.~\ref{fig:image3.png}(a) shows a snapshot detailing that ABP
accumulation predominantly occurs in regions of higher curvature. This
suggests that both the accumulation and the resulting bent conformation of
the polymer are more favorable at higher $\mathrm{Pe}$. Another interesting
observation is the pairing of the arms of the star polymer at high activity,
enabling it to adopt a conformation similar to that of a linear polymer.

At high activity, the accumulation of active particles facilitates the
formation of an asymmetric polymer structure---specifically, the pairing
of two arms while the third arm remains separate. To understand the
relationship between the adopted conformation and ABP accumulation, we
calculated the effective interaction energy between two arms of the star
polymer (details provided in the App.~\ref{appa}), relative to the conformation
with $\mathrm{R}_{A_i\mathrm{-}A_j}=9.0$, the value corresponding to their
separation at $\mathrm{Pe}=0$, as a function of $\mathrm{R}_{A_i\mathrm{-}
A_j}$ for various values of $\mathrm{Pe}$. In Fig.~\ref{fig:image3.png}(b),
in absence of activity with $\mathrm{Pe}=0$, the energy profile remains nearly
flat around zero, except for a positive deviation at around $\mathrm{R}_{A_i
\mathrm{-}j}=1.0$, which arises due to the self-avoidance of the polymer
beads. At finite activity with $\mathrm{Pe}=30$, the energy profile indicates
that conformations with smaller separations around $\mathrm{R}_{A_i\mathrm{-}
A_j}=1.0$ are less stable than those with larger separations. Interestingly,
at relatively high activity with $\mathrm{Pe}=50$, the energy profile shows a
stabilization of conformations with smaller distance around $\mathrm{R}_{A_i
\mathrm{-}A_j}\approx3$, supporting the tendency of the arms to pair at high
activity. This stabilization of conformations, such that the separation between
arms is minimized, goes hand in hand with the significant accumulation of ABPs
at such high activity levels. Therefore, we conclude that the stabilization of
paired-arm conformations is facilitated by the asymmetric accumulation and the
resulting exerted pressure of active particles on the polymer arms. Such effective attractions between passive tracers in an active bath have been previously reported. For example, when two passive particles are immersed in a bacterial bath, a short-range “attraction” emerges between them as the level of activity in the bath increases \cite{angelani2011effective}. Similar short-range attractions have also been observed for a linear polymer chain and a rigid rod in a bath of ABPs \cite{ni2015tunable, gandikota2022effective}.

In all cases shown in Fig.~\ref{fig:image3.png}(b), the depletion potential
$\mathrm{U}_{\mathrm{dep}}(\mathrm{R}_{A_i\mathrm{-}A_j})$ shows a significant
positive value at $\mathrm{R}_{A_i\mathrm{-}A_j}=1.0$, which increases with
growing $\mathrm{Pe}$. This confirms that the high positive value of $\mathrm{
U}_{\mathrm{dep}}(\mathrm{R}_{A_i\mathrm{-}A_j})$ at short distances arises due
to both polymer self-avoidance and the limited presence of ABPs between the arms,
a direct consequence of the star polymer geometry.

The inclusion of rigidity in a polymer exhibits more interesting features in
an active bath. For instance, a semiflexible linear polymer ($\kappa>0$)
adopts a hairpin-like structure due to the accumulation of active particles
in regions of high curvature and rigidity---a configuration not typically
observed in flexible polymers \cite{harder2014activity,shin2015facilitation,
goswami2022reconfiguration}.

In Fig.~\ref{fig:images2.png}(a) and (b) in App.~\ref{appb}, we plot the
probability density function for $\mathrm{R}_{A_i\mathrm{-}A_j}$ for flexible
($\kappa=0$) and semiflexible ($\kappa=75$) star polymers. As shown in
Fig.~\ref{fig:images2.png}(a), for the flexible polymer in the range $\mathrm{
Pe}=0\ldots20$, there is no emergence of a peak at lower values of $\mathrm{R}
_{A_i\mathrm{-}A_j}$; instead, the distribution broadens with increasing values
of $\mathrm{Pe}$. In contrast, for the semiflexible polymer case shown in
Fig.~\ref{fig:images2.png}(b), at $\mathrm{Pe}=0$, the distribution is peaked
at a higher value of $\mathrm{R}_{A_i\mathrm{-}A_j}$. On increasing the
activity to $\mathrm{Pe}=10$, a new peak emerges at a lower value of $\mathrm{
R}_{A_i\mathrm{-}A_j}$, and its intensity increases further at $\mathrm{Pe}=20$,
accompanied by a broader distribution at larger values of $\mathrm{R}_{A_i
\mathrm{-}A_j}$. A similar distribution of conformations is also observed for
flexible polymers at $\mathrm{Pe}=50$ in Fig.~\ref{fig:image2.png}, further
supporting the observation that higher bath activity reduces the effects of
polymer stiffness. As discussed previously, the shown distribution corresponds
to conformations in which two arms of the polymer are paired, while the third
arm is located at a certain distance. Based on these observations, we conclude
that the interplay between polymer rigidity and interactions with active
particles favors such conformations even at low $\mathrm{Pe}$ in the case of
semiflexible polymers.

In addition to calculating the inter-arm distance $\mathrm{R}_{A_i\mathrm{-}
A_j}$ and the arm-arm angle $\theta_{A_i\mathrm{-}A_j}$, we compute the
gyration tensor for the star polymers, defined as
\begin{widetext}
\begin{equation}
\label{shapeten}
S=\frac{1}{N_{\mathrm{beads}}}\begin{pmatrix}\sum\limits_i(x_i-x_{\mathrm{
com}})^2-\lambda_1^2&\sum\limits_i(x_i-x_{\mathrm{com}})(y_i-y_{\mathrm{com}})\\
\sum\limits_i(x_i-x_{\mathrm{com}})(y_i-y_{\mathrm{com}})&\sum\limits_i(y_i-
y_{\mathrm{com}})^2-\lambda_2^2\end{pmatrix},
\end{equation}
\end{widetext}
where $x_{\mathrm{com}}$ and $y_{\mathrm{com}}$ denote the coordinates of the
polymer's center of mass, and $N_{\mathrm{beads}}=61$ is the total number of
beads in the star polymer. To analyze the shape anisotropy, we diagonalize the
gyration tensor (\ref{shapeten}) to obtain its principal components
\begin{equation}
S=\mathrm{diag}(\lambda_1^2,\lambda_2^2),
\end{equation}
where $\lambda_1^2$ and $\lambda_2^2$ are the eigenvalues of the tensor. The
first invariant of $S$ yields the squared radius of gyration
\begin{equation}
\mathrm{R}_g^2=\mathrm{Tr}(S)=\lambda_1^2+\lambda_2^2.
\end{equation}
We compute the time and ensemble-averaged value of $R_g^2$ as function of the
P{\'e}clet number $\mathrm{Pe}$ across multiple trajectories. Specifically, in
Fig.~\ref{fig:images2.png}(c) in App.~\ref{appb}, we plot the averaged squared
radius of gyration $\langle\mathrm{R}_g^2(\mathrm{Pe})\rangle$ as a function of
the bath activity for both flexible (black) and semiflexible (red) star polymers.
At vanishing activity ($\mathrm{Pe}=0$), the flexible star polymer has compact,
coiled conformations, resulting in smaller $\langle\mathrm{R}_g^2\rangle$ values.
As the activity increases, collisions with active particles lead to a swelling
of the polymer, reflected in increased $\langle\mathrm{R}_g^2\rangle$ values.

In contrast, the semiflexible star polymer adopts an extended conformation
at $\mathrm{Pe}=0$, primarily due to the effect of arm rigidity. As activity
increases, $\langle\mathrm{R}_g^2\rangle$ decreases, which we attribute to
the pairing and bending of arms caused by the asymmetric accumulation of
active particles in regions of high curvature of the polymer. An intriguing
feature emerges at high activity levels: the effect of activity starts to
dominate over the rigidity of the arms, as mentioned above. As a result, the
values of $\langle\mathrm{R}_g^2\rangle$ for both flexible and semiflexible
star polymers begin to converge, indicating that both types of polymers adopt
similar conformations under strong activity. This swelling and collapsing
behavior of the flexible and semiflexible star polymers in an active bath is
similar to that observed for flexible and semiflexible linear polymers,
respectively \cite{shin2015facilitation,harder2014activity}.

\subsubsection{Dynamics of the star polymer}

In the existing literature it has been explored how activity of bath
particles facilitates a prolonged persistent motion of asymmetric hard
passive particles \cite{kaiser2014transport,angelani2010geometrically,
smallenburg2015swim}. The persistent motion of an object in a given
direction in some sense closely mirrors what Feynman described in his
well-known "Ratchet-and-Paw" lecture, in which he explored the second law
of thermodynamics through the lens of statistical mechanics. The idea of
harnessing work from a chaotic environment through rectification depends on
breaking both time-reversal and spatial-inversion symmetries. In Feynman's
example, time-reversal symmetry is disrupted by using two thermal baths at
different temperatures, while in another case, it is broken by a bath of
active particles with inherently irreversible dynamics \cite{o2022time}.
This disruption violates detailed balance, a condition in equilibrium
systems where forward and reverse processes occur with equal likelihood.
When time-reversal symmetry is no longer valid, breaking spatial
symmetry---such as by using an asymmetrically shaped gear or shuttle---can
enable rectification and thus allow for directed motion or work extraction
\cite{curie1894symetrie,reimann2002brownian}. Experimentally, energy
extraction from a symmetric stochastic process in the sense of a Maxwell
demon could indeed be demonstrated for a colloidal particle \cite{yael_maxw}.

It is important to understand such partially-rectified dynamic in the context
of soft particles of elongated topology \cite{angelani2010geometrically}; here
some recent work focused on a soft linear polymer in an active bath where the
polymer showed pronouncedly enhanced dynamics \cite{shin2015facilitation}. To
understand the dynamics of a passive star polymer in an active bath, we now
explore the behavior of a three-armed flexible star polymer immersed in a bath
of ABPs as function of time. Due to the presence of the active particles, the
star polymer switches between several conformations, as discussed in the
previous section. As we show now this also effects intriguing dynamic behavior.

To quantify these effects, we track the center-of-mass position $\mathrm{r}
^{\mathrm{com}}_i$ of the star polymer and compute the component-wise
time-averaged mean-squared displacement (TAMSD) as function of the lag time
$\tau$ \cite{he,pt,pccp}
\begin{equation}
\label{tamsd}
\overline{\delta r_i^2(\tau)}=\frac{1}{T_{\mathrm{max}}-\tau}\int_0^{T_{
\mathrm{max}}-\tau}\Big(r^{\mathrm{com}}_i(t+\tau)-r^{\mathrm{com}}_i(t)
\Big)^2dt,
\end{equation}
whose average over a number $N$ of independent trajectories is the
ensemble-averaged (mean) TAMSD \cite{he,pt,pccp}
\begin{equation}
\label{eatamsd}
\left<\overline{\delta r_i^2(\tau)}\right>=\frac{1}{N}\sum_{i=1}^N\overline{
\delta r_i^2(\tau)}.
\end{equation}
For Brownian motion, the mean TAMSD $\left<\overline{\delta r_i^2(\tau)}
\right>$ grows linearly with lag time $\tau$. However, in the presence of
ergodic anomalous dynamics, we observe a power-law scaling of the form
\begin{equation}
\left<\overline{\delta r_i^2(\tau)}\right>\propto\tau^{\alpha},
\end{equation}
where $\alpha<1$ indicates subdiffusion and $\alpha>1$ indicates
superdiffusion \cite{pccp,report,franosch}.

\begin{figure*}
\includegraphics[width=0.70\linewidth]{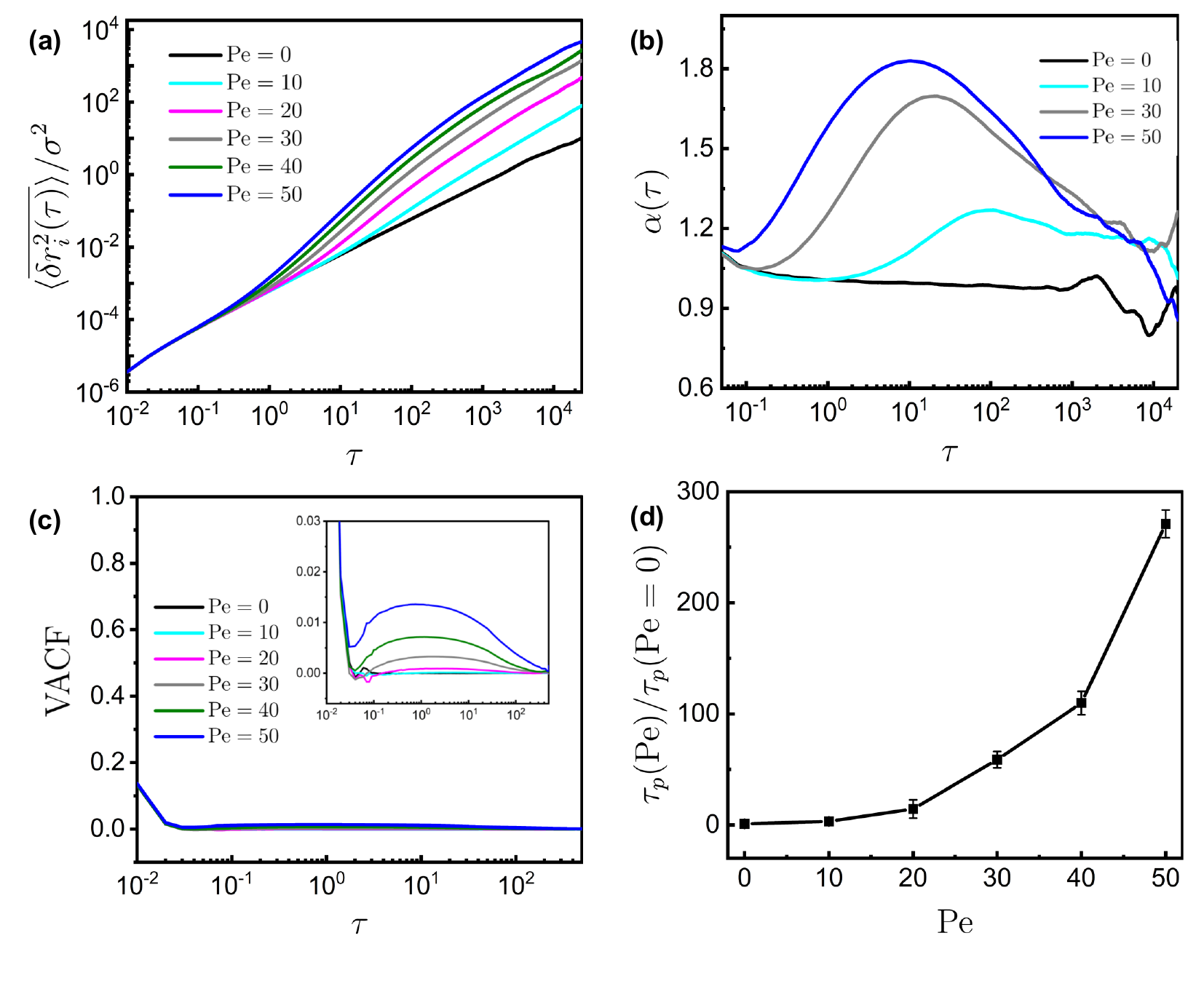}
\caption{ Log-log plot of the time evolution of (a) the mean TAMSD $\left<
\overline{\delta r_i^2(\tau)}\right>$; log-linear plots of (b) the time-local
anomalous diffusion exponent $\alpha(\tau)$, along with the (c) VACF of the
motion for different values of $\mathrm{Pe}$ and (d) the relative effective
persistence time $\tau_p(\mathrm{Pe})/\tau_p(\mathrm{Pe}=0)$ at some value
of $\mathrm{Pe}$ divided by its inactive value, as a function of $\mathrm{Pe}$
for the center of mass motion of the star polymer. Inset in panel (c) shows a magnified view of the same.}
\label{fig:image4.png}
\end{figure*}

In Fig.~\ref{fig:image4.png}(a), we plot the time evolution of the mean TAMSD
(\ref{eatamsd}) for different values of the P{\'e}clet number $\mathrm{Pe}$. It
is evident that upon increasing the activity $\mathrm{Pe}$ of the active
particles leads to an enhancement of the motion as measured by $\left<\overline{
\delta r_i^2(\tau)}\right>$. This enhancement arises from frequent encounters and
asymmetric accumulation of active particles around the polymer. The extent of
this enhancement is related to the persistent motion of the polymer, which
is evident in Fig.~\ref{fig:image4.png}(b), where we show how $\alpha(\tau)$
increases with $\mathrm{Pe}$, indicating a pronounced transient superdiffusive
behavior and thus enhanced persistence. Further evidence of the prolonged
persistent motion is provided by the VACF and the effective persistence time
$\tau_p$ of the center of mass motion of a passive star polymer.\footnote{This
persistence time is distinct from the persistence time $\tau_R$ of ABPs.} as
shown in Fig.~\ref{fig:image4.png}(c) and (d), respectively. The VACF is here
defined over the time average
\begin{equation}
\mathrm{VACF}=\frac{\left<\overline{v_i(t+\tau)v_i(t)}\right>}{\left<
\overline{v_i^2(t)}\right>},
\end{equation}
where $v_i$ is a component of the velocity $\mathbf{v}_{\mathrm{com}}$ of
the polymer's center of mass. The effective persistence time ($\tau_p$) is
defined as the duration during which the center of mass of the polymer
maintains its motion in a given direction. It is mathematically expressed as
\begin{equation}
\label{taup}
\tau_p=\int_0^{t^*}\mathrm{VACF}d\tau,
\end{equation}
where $t^*$ is the time at which the VACF reaches zero.

In Fig.~\ref{fig:image4.png}(c), we observe that beyond $\mathrm{Pe}=20$
the VACF exhibits a pronounced positive temporal correlation that extends
over a longer period for the higher $\mathrm{Pe}$ values $\mathrm{Pe}=30$
and $50$, indicating prolonged persistent motion. To quantify the enhancement
in the persistence, we plot the relative persistence time (\ref{taup}) as a
function of $\mathrm{Pe}$ in Fig.~\ref{fig:image4.png}(d). For $\mathrm{Pe}
\leq20$, there is no significant effect, and the ratio remains nearly constant.
However, beyond $\mathrm{Pe}\approx20$, there is a noticeable increase of the
persistence time with activity. Notably, the $\mathrm{Pe}$ value at which this
significant increase in the persistence time occurs coincides with the
emergence of a distinct polymer conformation: the pairing of two arms while
the third arm remains separated. We conclude that the persistent motion of the
star polymer is closely linked to this conformation, which is governed by the
asymmetric accumulation of active particles.

\begin{figure*}
\includegraphics[width=0.70\linewidth]{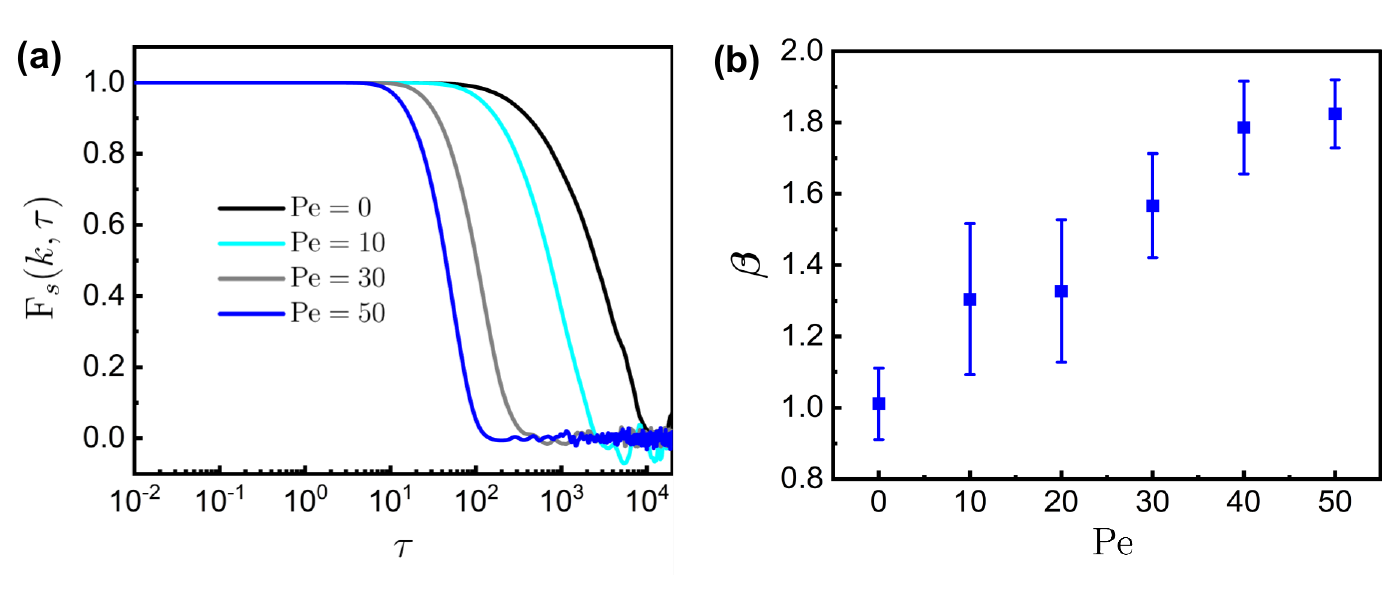}
\caption{Log-linear plots showing the time evolution of the intermediate
scattering function (\ref{intscat}) for (a) wave number $k=1$. Panel
(b) shows the change in the compressed exponential coefficient $\beta$
defined in Eq.~(\ref{eq:fitting2}) as a function of $\mathrm{Pe}$.}
\label{fig:image5.png}
\end{figure*}

Another fundamental quantity of interest for understanding the star dynamics is
the intermediate scattering function defined as function of the difference in
the center of mass position as separated by the lag time $\tau$,
\begin{equation}
\label{intscat}
\mathrm{F}_{s}({k},\tau)=\left<\exp\Big(-ik(r_j^{\mathrm{com}}(
\tau)-r_j^{\mathrm{com}}(0))\Big)\right>,
\end{equation}
where $r_j^{\mathrm{com}}(\tau)$ and $r_j^{\mathrm{com}}(0)$ are the positions
of the center of mass of the polymer at times $\tau$ and $0$, respectively. The
symbol $\langle\cdot\rangle$ again denotes averaging over different trajectories.
In Fig.~\ref{fig:image5.png}(a), we plot the time evolution of $\mathrm{F}_s
(k,\tau)$ for $k=1.0$ at different values of $\mathrm{Pe}$.
It is evident that the decay of $\mathrm{F}_{s}({k},\tau)$ becomes faster
with increasing $\mathrm{Pe}$ as compared to the case $\mathrm{Pe}=0$ in absence
of activity.

In the case of Brownian dynamics, the dependence of the intermediate scattering
function (\ref{intscat}) on $k$ is given by
\begin{equation}
\mathrm{F}_{s}({k},\tau)=\exp\Big(-Dk^2\tau\Big)=
\exp\left(-k^2\left[\frac{\tau}{\tau_s}\right]\right),
\label{eq:fitting1}
\end{equation}
where in the second equality we used the fact that the diffusivity $D$ is
related to the relaxation time $\tau_s$ via $\tau_s=1/D$ \cite{clark1970study}.

Since our star polymers in the active bath exhibit pronounced persistent
motion---especially at high activity---deviations from the simple exponential
form (\ref{eq:fitting1}) are expected. To account for this, we employ the
modified exponential form
\begin{equation}
\mathrm{F}_{s}({k},\tau)=\exp\left(-\left[\frac{\tau}{\tau_s(k)}
\right]^{\beta}\right),
\label{eq:fitting2}
\end{equation}
where $\beta$ is the stretching coefficient. For $0<\beta<1$ the form
(\ref{eq:fitting2}) is referred to as a "stretched exponential function"
\cite{rama} or Kohlrausch-Williams-Watts (KWW) form, historically linked to
the works of Kohlrausch \cite{kohlrausch} and Williams and Watts
\cite{watts}; see also \cite{vlad,huber,mike,palmer}.
Values of $\beta>1$ in relation (\ref{eq:fitting2}) produce a "compressed
exponential" behavior. The exponent $\beta$ characterizes the extent to
which the decay of $\mathrm{F}_{s}({k},\tau)$ is faster than the simple
exponential decay, the latter being the hallmark of Brownian dynamics. In
Fig.~\ref{fig:image5.png}(b), we plot the variation of $\beta$ with $\mathrm{
Pe}$ for $k=1$. At $\mathrm{Pe}=0$, we observe $\beta=1$, and
Eq.~\ref{eq:fitting2} reduces to the Brownian form given in
Eq.~\ref{eq:fitting1}. As activity increases, $\beta$ becomes greater than
unity and continues to grow with increasing $\mathrm{Pe}$. A faster relaxation
dynamics with $\beta>1$ is in fact typical in systems exhibiting long
persistence times \cite{szamel2024extremely,tjhung2020analogies}.
We show that the intermediate scattering function relaxes more rapidly at
larger values of $k$ at other activity levels, as well, see
Fig. \ref{fig:images3} in App.~\ref{appc}.

\subsection{Comparison of the behavior of a passive particle and a star polymer
in a dilute active bath}

\begin{figure*}
\includegraphics[width=0.98\linewidth]{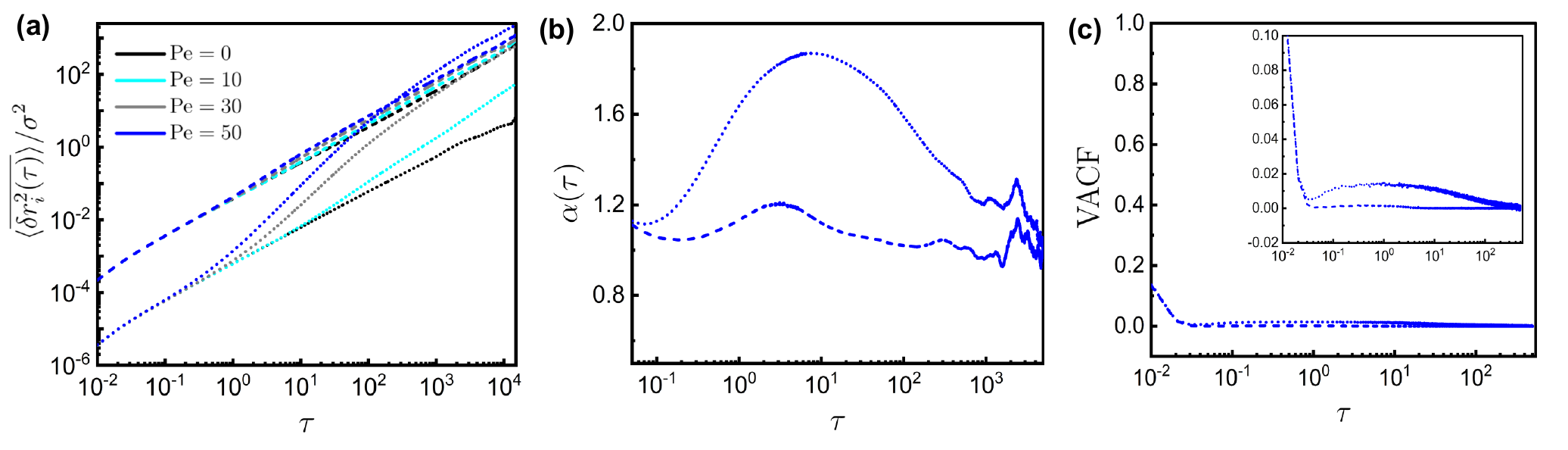}
\caption{Time evolution of the (a) mean TAMSD $\left<\overline{\delta^2
r_i(\tau)}\right>$ on log-log scale for different activities $\mathrm{Pe}$,
(b) time-local anomalous diffusion exponent $\alpha(\tau)$ on log-linear
scale at $\mathrm{Pe}=50$, and (c) VACF on log-linear scale at $\mathrm{Pe}=
50$ for the center of mass (COM) of the star polymer (dotted lines) and the
single colloidal particle (dashed lines). The inset in panel (c) shows the
significant range of positive correlations as compared to the single particle,
whose VACF remains approximately zero throughout.}
\label{fig:image6.png}
\end{figure*}

The dynamics of passive colloidal particles and linear polymers in an active
bath were shown to enhance the particle diffusion, often effecting (transient)
superdiffusion \cite{pietzonka2017entropy,chaki2018entropy,chaki2019effects,
chaki2019enhanced,ye2020active,shea2022passive,goswami2023trapped,
grossmann2024non,anand2020conformation,shin2015facilitation,kaiser2015does,
chaki2019enhanced,goswami2022reconfiguration,samanta2016chain,
aporvari2020anisotropic,ghosh2014dynamics,osmanovic2017dynamics}.
In particular, it was shown that linear polymers in active baths have
intriguing conformational and dynamic properties. To further investigate the
effect of the polymer topology, we compare the dynamics of our star polymer
with other entities, starting with a single particle. To this end we simulated
a passive particle of diameter $\sigma$ in a dilute active bath, the results
are displayed in Fig.~\ref{fig:image6.png}, in comparison with the dynamics of
our three-armed star polymer. In Fig.~\ref{fig:image6.png}(a) we show the
component-wise mean TAMSDs of the single particle position $r_i$ and of the
center-of-mass position $r^{\mathrm{com}}_i$ of the star polymer. In the
passive case ($\mathrm{Pe}=0$), the dynamics remain normal-diffusive,
regardless of the tracer topology, as shown in Fig.~\ref{fig:image6.png}(a).
The differences in the amplitude arise due to the effective tracer size: the
colloidal particle exhibits a faster dynamics in comparison to the larger star
polymer. With increasing activity, the dynamics of colloidal particles show
only a slight change, indicating that activity has no profound effect on their
motion. In contrast, the star polymer displays a significant enhancement in the
mean TAMSD with increasing activity. This enhancement is also evident from the
larger values of the anomalous diffusion exponent $\alpha(\tau)$ for the case
$\mathrm{Pe}=50$ for the star polymer as compared to the passive colloidal
particle in Fig.~\ref{fig:image6.png}(b).

\begin{figure*}
\includegraphics[width=0.98\linewidth]{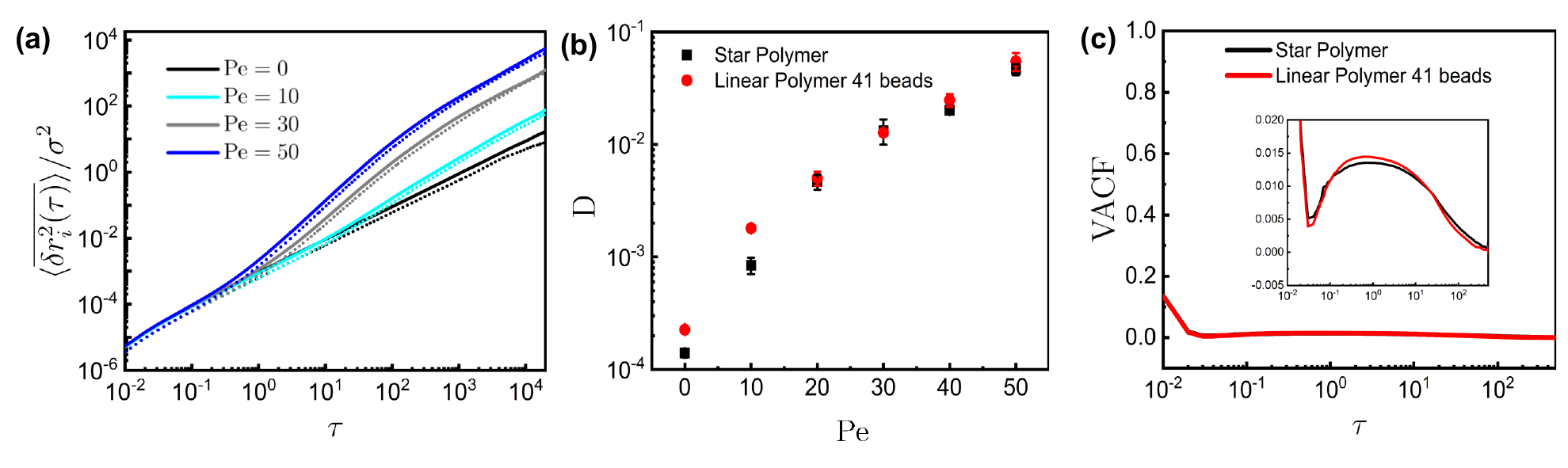}
\caption{(a) Log-log plot of the mean TAMSD versus the lag time for the center
of mass (COM) of the star polymer (dotted lines) and a linear polymer (41 beads,
solid lines). (b) Linear-log plot of the diffusivity of the COM for the star
polymer (black) and the linear polymer (red) as a function of $\mathrm{Pe}$.
(c) VACF on a log-linear scale at $\mathrm{Pe}=50$ for the COM of the star
polymer (61 beads black) and the linear polymer (41 beads, red). The inset in
(c) shows a magnified view with a clear local minimum at short times.}
\label{fig:image7}
\end{figure*}

The high values of $\alpha(\tau)$ along with the longer relaxation time
toward the normal-diffusive regime with $\alpha=1$ observed for the star
polymer, as compared to the colloidal particle, indicate a more persistent
and effective motion in the presence of activity. The prolonged persistent
motion of the star polymer with a positive VACF, relative to the passive
colloidal particle, as shown in Fig.~\ref{fig:image6.png}(c) supports the
increase of the mean TAMSD for larger P{\'e}clet numbers. Note that for
the highest P{\'e}clet number shown ($\mathrm{Pe}=50$), the mean TAMSD of
the star polymer even overshoots the values of the single particle at
longer times, despite the larger size of the star polymers. The VACF shown
in Fig.~\ref{fig:image6.png}(c) demonstrates the relatively long time range
over which the star polymer has significant positive correlations, before
eventually decaying to zero.

\subsection{Comparison of the behavior of a linear polymer with 41 beads
and the three-armed star polymer in a dilute active bath}

We now compare the dynamics of our star polymer with that of a linear polymer.
We choose the size of 41 beads for the linear polymer, as in this case we
expect that the behavior of star and linear polymers become approximately the
same at high activity, when the star polymer has two aligned arms. The mean
TAMSD of both polymers is shown in Fig.~\ref{fig:image7}(a) along with the
long-time diffusivity
\begin{equation}
D=\lim_{\tau\to\infty}\frac{1}{4\tau}\left<\overline{\delta r_i^2(\tau)}\right>
\end{equation}
in Fig.~\ref{fig:image7}(b). Indeed, at high $\mathrm{Pe}$ the effective
long-time diffusivity of both topologies converge, despite the fact that the
star polymer consists of 61 monomers as compared to the 41 of the linear chain.
This difference causes the smaller diffusivity and amplitude of the mean TAMSD
at smaller $\mathrm{Pe}$. In Fig.~\ref{fig:image7}(c), we plot the VACF of the
center of mass for the linear polymer (solid line) and the star polymer (dotted
line) for $\mathrm{Pe}=50$. In this high-activity case both show very similar
behavior with extended-in-time positive correlations.

\begin{figure*}
\includegraphics[width=0.98\linewidth]{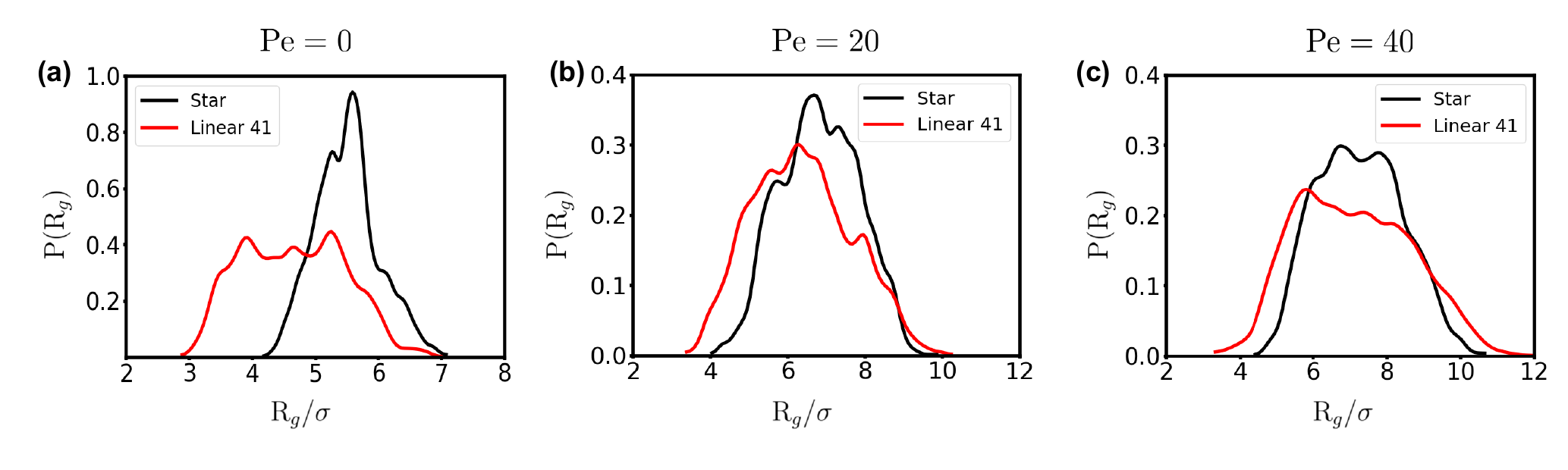}
\caption{Probability density function of the gyration radius for the
star polymer (61 beads, black) and the linear polymer (41 beads, red) for (a)
$\mathrm{Pe}=0$, (b) $\mathrm{Pe}=20$, and (c) $\mathrm{Pe}=40$.}
\label{fig:image8}
\end{figure*}

\begin{figure*}
\includegraphics[width=0.98\linewidth]{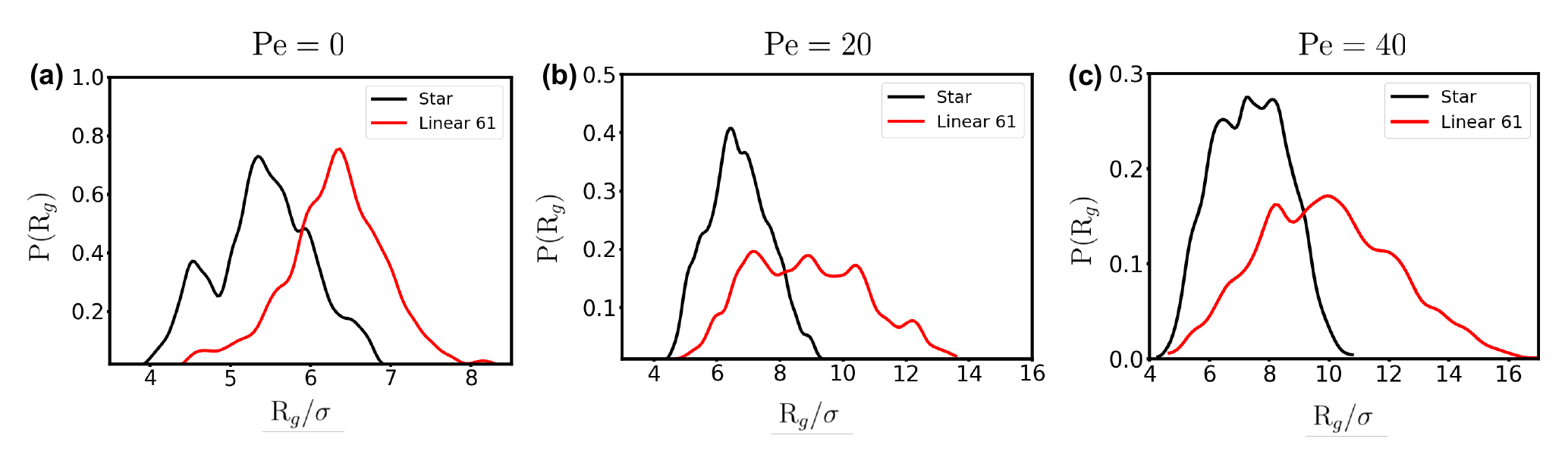}
\caption{Probability density function of the gyration radius for the
star polymer (black line) and the linear polymer (61 beads, red line) at (a)
$\mathrm{Pe}=0$, (b) $\mathrm{Pe}=20$, and (c) $\mathrm{Pe}=40$.}
\label{fig:image9.png}
\end{figure*}

To understand the structural features of the polymers underlying the observed
convergence of their dynamics at high $\mathrm{Pe}$, we plot the probability
density function of the radius of gyration $\mathrm{R}_g$ for the
star polymer (black line) and the linear polymer (red line) at different values
of the P{\'e}clet number in Fig.~\ref{fig:image8}. Specifically, in absence of
activity ($\mathrm{Pe}=0$) in Fig.~\ref{fig:image8}(a), the density for the
star polymer is peaked at a higher value of $R_g$ compared to that of the
linear polymer. Additionally, the distribution for the linear polymer is
more broadly spread. Upon increasing the activity to $\mathrm{Pe}=20$, the
difference between the distributions diminishes, and at high activity with 
$\mathrm{Pe}=40$ in Fig.~\ref{fig:image8}(c), the distributions cover almost
the same range of $\mathrm{R}_g$. The convergence in the distributions
for the linear and star polymers at high activity confirms the similarity in
their structural features. This also supports the pairing of two arms of the
star polymer, while the third arm remains at a variable distance, resulting
in a conformation similar to that of a linear polymer in an active bath,
due to the effective pressure exerted by the accumulating ABPs in regions of
higher curvature. In both cases, the accumulation of ABPs increases with
increasing activity, as demonstrated in Fig.~\ref{fig:images4.png} in
App.~\ref{appd}. The similarity in the structure of linear and star polymers
at high $\mathrm{Pe}$ is also evident in Fig.~\ref{fig:images5.png}(a) in
App.~\ref{appe}, where the difference in the squared radius of gyration
between the star and linear polymers decreases with increasing $\mathrm{Pe}$
and nearly vanishes at high $\mathrm{Pe}$.

\subsection{Comparison of the behavior of a linear polymer with 61 beads
and a three-armed star polymer in a dilute active bath}

\begin{figure*}
\includegraphics[width=0.98\linewidth]{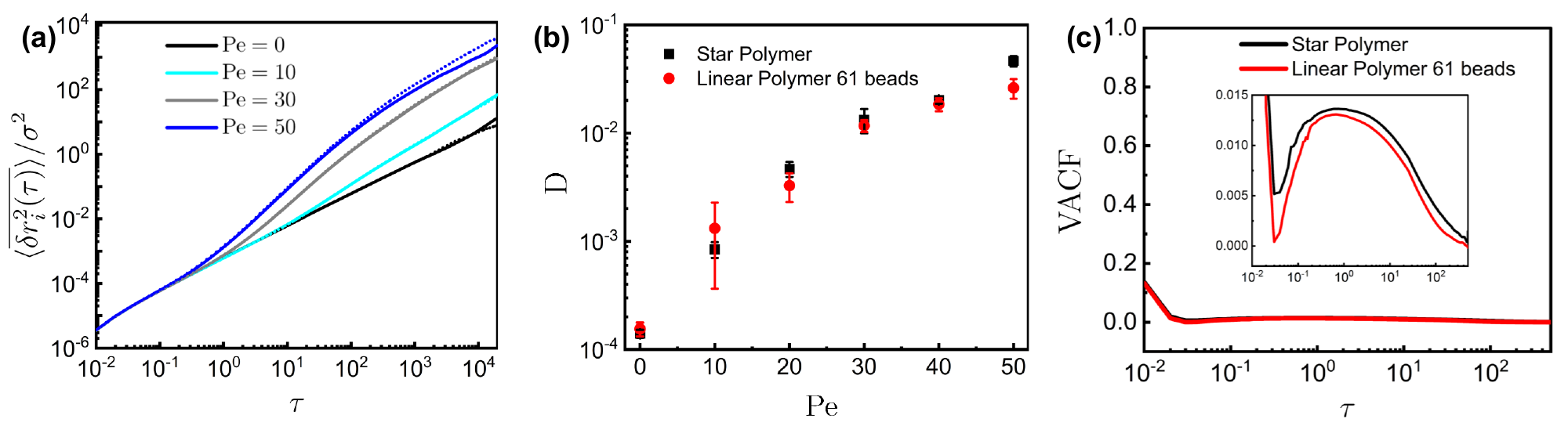}
\caption{(a) Log-log plot of the mean TAMSD versus lag time for the center of
mass (COM) motion of the linear polymer (61 beads, solid lines) and the star
polymer (dotted lines). (b) Linear-log plot of the diffusivity of the COM for
the star polymer (black) and the linear polymer (red) as a function of
$\mathrm{Pe}$. (c) VACF on a log-linear scale at $\mathrm{Pe}=50$ for the
center of mass (COM) motion of the star polymer (black) and the linear polymer
(red). The inset in (c) shows a magnified view.}
\label{fig:image10.png}
\end{figure*}

In Fig.~\ref{fig:image9.png}, we compare the radius of gyration of a
three-armed star polymer and a linear polymer for the case when both have the
same total number of monomers. In Fig.~\ref{fig:image9.png}(a), we plot the
probability density function of the radius of gyration for the
three-armed star polymer (black) and the linear polymer (red) at $\mathrm{Pe}
=0$. The distribution is peaked at a higher value for the linear polymer as
compared to the star polymer, reflecting the effect of topology on the polymer
extension. As the activity increases to $\mathrm{Pe}=20$, shown in
Fig.~\ref{fig:image9.png}(b), and then to $\mathrm{Pe}=40$ shown in
Fig.~\ref{fig:image9.png}(c), the difference between the two topologies
becomes more pronounced; specifically, the distribution for the linear polymer
broadens and shifts further to higher values of $\mathrm{R}_g$. This difference
in extension is also evident in Fig.~\ref{fig:images5.png} in App.~\ref{appe},
where we plot the difference in the average squared radius of gyration between
the linear and star polymers as a function of activity. The difference indeed
increases monotonically with $\mathrm{Pe}$, supporting the intuition that the
pairing of arms in the star polymer, along with its more complex topology,
limits its extension under active conditions.

To analyze further the effect of topology on the polymer dynamics, we plot the
time evolution of the mean TAMSD of the center of mass and the corresponding
long-time diffusivity in Fig.~\ref{fig:image10.png}. Fig.~\ref{fig:image10.png}(a)
demonstrates that at lower activity levels ($\mathrm{Pe}\leq20$), the mean
TAMSD of the center of mass for both the linear polymer (61 beads) and the star
polymer with three arms (also 61 beads) are nearly identical. However, at higher
activity, specifically at $\mathrm{Pe}=50$, the mean TAMSD of the star polymer
exceeds that of the linear polymer. This supports the observation related to
the structural features of the polymers: while the linear polymer extends
more readily, the arms of the star polymer tend to pair, limiting its extension.
A similar trend is observed in Fig.~\ref{fig:image10.png}(b), where the
diffusivity of the star polymer is shown to increase significantly at high
$\mathrm{Pe}$ as compared to that of the linear polymer. This enhancement in
the dynamics of the star polymer relative to the linear polymer at high activity
is further supported by the time evolution of the VACF, shown in
Fig.~\ref{fig:image10.png}(c). Namely, the VACF for the star polymer (dotted
line) is more positive than that of the linear polymer (solid line), confirming
the more persistent motion of the star polymer in this high activity regime.

\section{Conclusion}
\label{sec:conclusion}

We presented an investigation of the dynamics and conformational changes of
star polymers in a dilute bath of ABPs using two-dimensional molecular dynamics
simulations, and compared their behavior with tracers of other topologies. Our
results reveal distinctive conformational responses of star polymers in the
active bath. Notably, beyond a certain threshold of activity as measured by
the P{\'e}clet number $\mathrm{Pe}$, the star polymer adopts a configuration
in which two arms pair while the third remains extended---an arrangement
driven by the asymmetric accumulation of active particles in regions of higher
chain curvature. The pairing of the arms of the star polymer in the active
bath resembles observations from previous studies, in which passive tracers
(both linear and colloidal) experience depletion-induced attractive forces due
to the accumulation of active particles \cite{gandikota2022effective,
angelani2011effective,ni2015tunable}. This asymmetry, combined with the arm
pairing of the flexible star polymers, results in prolonged directed motion,
as evidenced by the enhanced mean TAMSD, sustained positive VACF, and an
increased persistence time with increasing $\mathrm{Pe}$.

Semiflexible star polymers exhibit this arm-pairing conformation at lower
activity levels as compared to their flexible counterparts, suggesting that
polymer rigidity promotes such structural asymmetry. When compared to passive
colloidal particles---each having the same size as a single polymer bead---the
star polymer exhibits faster and more persistent center-of-mass dynamics at
high activity, owing to the same mechanisms of arm pairing and active particle
accumulation. This difference is clearly reflected in the higher mean TAMSD
larger time-local anomalous diffusion exponents, and more pronounced positive
VACF observed for the star polymer as compared to the colloidal particles at
high $\mathrm{Pe}$. The persistent dynamics of the star polymer arise from its
asymmetric structure and the accumulation of active particles in regions of
high curvature \cite{shin2015facilitation,di2010bacterial,
angelani2010geometrically}.

Interestingly, at high activity, the conformation of the star polymer---with
two arms paired and one extended---closely resembles that of a linear
polymer. This structural similarity is reflected in their dynamics as well. To
explore this, we compared the dynamics of a three-armed star polymer (61 beads)
with those of a linear polymer consisting of 41 beads. The results show that
their dynamics indeed converge at high $\mathrm{Pe}$.

Furthermore, when comparing a star and a linear polymer with the same number
of beads, the star polymer still demonstrates faster and more persistent
dynamics at high activity. This distinction is attributed to their
conformational differences, as indicated by the smaller radius of gyration
of the star polymer relative to the linear chain. Additionally, the velocity
autocorrelation function confirms that the star polymer retains a higher
level of temporal correlation than the linear polymer under similar conditions.

We are hopeful that, apart from the interesting physical observations, this
study will also help guiding the design of polymer-based drug carriers for
efficient induced cargo transport in complex, non-equilibrium media, ultimately
having in mind targeted drug delivery purposes.

\begin{acknowledgments}
R.S.Y. thanks IIT Bombay for the fellowship. R.C. acknowledges IRCC-IIT Bombay
(project no. RD/0518-IRCCAW0-001) for funding. We acknowledge the SpaceTime-2
supercomputing facility at IIT Bombay for the computing time. R.M. acknowledges
the computer facility at the University of Potsdam, Germany and the German
Science Foundation (DFG, grant ME 1535/16-1 and ME 1535/22-1).
\end{acknowledgments}

\appendix

\section{Calculation of the effective energy between arms pair of star polymer}
\label{appa}

\begin{figure}
\includegraphics[width=0.50\linewidth]{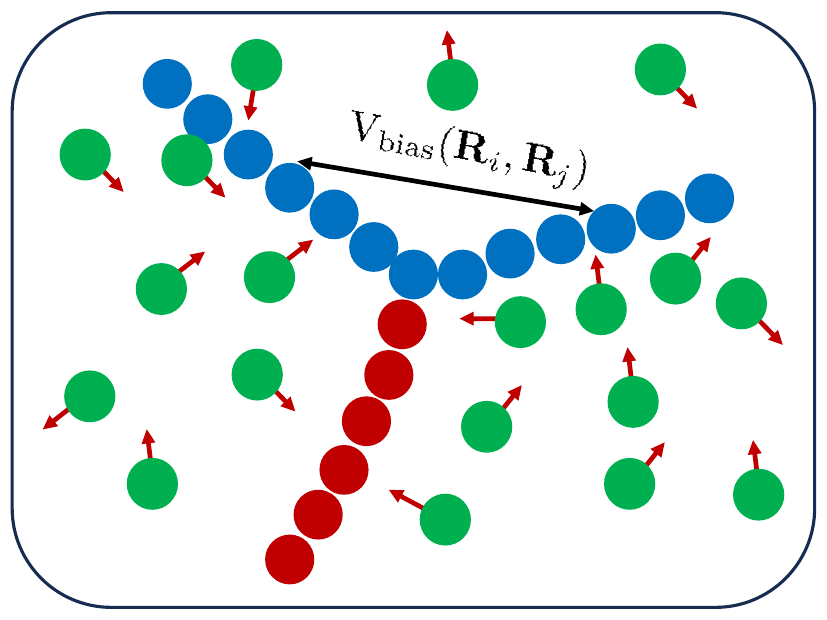}
\caption{Schematic depiction of the model system (not to scale): a single
star polymer with three arms immersed in an active bath. This system is
simulated separately under the influence of a harmonic bias potential,
$\mathrm{V}_{\mathrm{bias}}(\mathbf{R}_i,\mathbf{R}_j)$, applied between
pairs of arms (shown in blue), and immersed in a dilute bath of ABPs
(shown in green). The red arrows indicate the instantaneous directions of
the ABPs.}
\label{fig:images1.png}
\end{figure}

To obtain the effective interaction energy between a pair of arms of the star
polymer placed in a dilute active bath at a given center-of-mass separation
$\mathrm{R}_{A_i\mathrm{-}A_j}$, we enforce a specific separation by applying a
harmonic bias potential. The star polymer is immersed in an active bath, as
schematically illustrated in Fig.~\ref{fig:images1.png}, where the polymer
beads are shown in blue and the active particles in green. The harmonic bias
potential used to maintain the specified arm separation is given by
\cite{gandikota2022effective,ni2015tunable,kawaguchi2013molecular}
\begin{equation}
\mathrm{V_{\mathrm{bias}}}(\mathbf{R}_i,\mathbf{R}_j)=\frac{1}{2}k'(|\mathbf{
R}_i-\mathbf{R}_j|-\mathrm{R}_0)^2,
\end{equation}
where $\mathbf{R}_i$ and $\mathbf{R}_j$ are the center-of-mass position vectors
of the two polymers, $k'$ is the harmonic force constant, $\mathrm{R}_0$ is
the equilibrium (or target) separation, and $\mathrm{R}_{A_i\mathrm{-}Aj}=
|\mathbf{R}_i-\mathbf{R}_j|$ is the Euclidean distance between the centers of
mass. We start with a separation $\mathrm{R}_{A_i\mathrm{-}A_j}=9.0$, which
corresponds to the distance between the selected pair of arms at $\mathrm{Pe}
=0$. The star polymer is then simulated in an active bath for the specified
value of $\mathrm{R}_{A_i\mathrm{-}A_j}$ under the influence of the bias
potential until the system reaches a steady state.

Using the steady-state configurations, we calculate the effective force
$\mathrm{F}^S(\mathrm{R}_{A_i\mathrm{-}A_j})$ between two arms of the
star polymer at a center-of-mass separation $\mathrm{R}_{A_i\mathrm{-}A_j}$.
Since the calculated force $\mathrm{F}^S(\mathrm{R}_{A_i\mathrm{-}A_j})$
includes contributions from the bias potential, the actual force arising from
the accumulation of active particles is obtained by subtracting the bias force
from the measured $\mathrm{F}^S(\mathrm{R}_{A_i\mathrm{-}A_j})$. The
corresponding potential is then calculated by integrating the net force
$\mathrm{F}^{S_{\mathrm{net}}}(\mathrm{R}_{A_i\mathrm{-}A_j})$ over the
separation distance,
\[
U_{\mathrm{dep}}(\mathrm{R}_{A_i\mathrm{-}A_j})=-\int\mathrm{F}^{S_{\mathrm{
net}}}(\mathrm{R}_{A_i\mathrm{-}A_j})dr,
\]
as described in \cite{chakrabarti2006short}. The results are first averaged
over time after the system reaches a steady state, and then further averaged
over 10 different trajectories.

\section{Comparison of the structural features of flexible and semiflexible
star polymers in an active bath}
\label{appb}

\begin{figure*}
\includegraphics[width=0.98\linewidth]{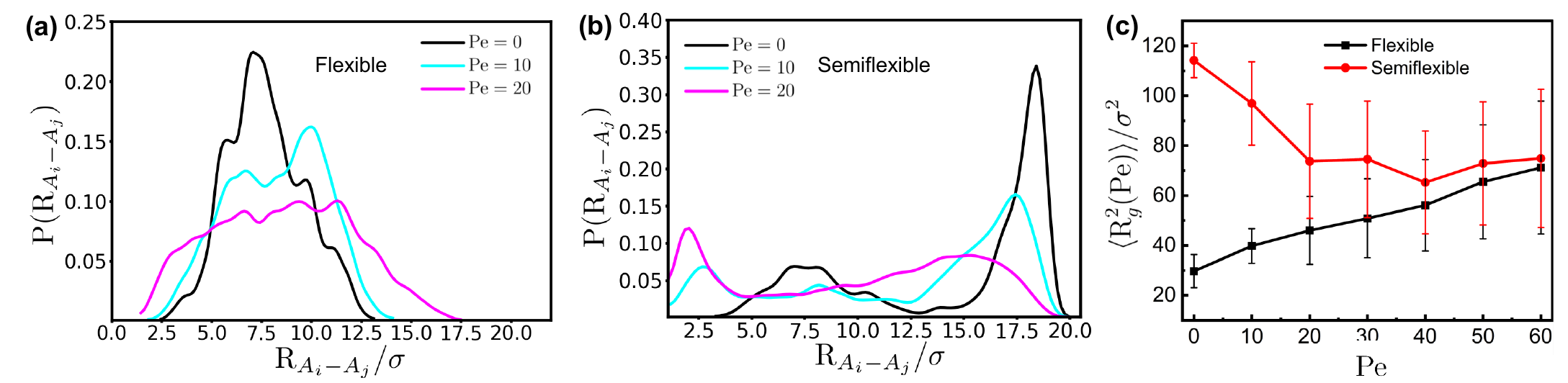}
\caption{Probability density function of $\mathrm{R}_{A_i\mathrm{-}A_j}^2$
for (a) flexible and (b) semiflexible star polymers. (c) Change in the
averaged squared radius of gyration, $\langle\mathrm{R}_g^2\rangle$ as
function of $\mathrm{Pe}$ for flexible (black) and semiflexible (red) star
polymers.}
\label{fig:images2.png}
\end{figure*}

To understand the effect of polymer rigidity on the structural features of the
star polymer, we plot the probability density function of the arm-arm
separation $\mathrm{R}_{A_i\mathrm{-}A_j}$ for the flexible star polymer in
Fig.~\ref{fig:images2.png}(a) and the semiflexible star polymer in
Fig.~\ref{fig:images2.png}(b). Additionally, we plot the average squared
radius of gyration as function of $\mathrm{Pe}$ in Fig.~\ref{fig:images2.png}(c).
The black line corresponds to the flexible polymer ($\kappa=0$), and the red line
depicts the semiflexible polymer ($\kappa=75$).

\section{Intermediate scattering function of the center of mass motion of
the star polymer for different wave numbers in a dilute active bath}
\label{appc}

\begin{figure}
\includegraphics[width=0.8\linewidth]{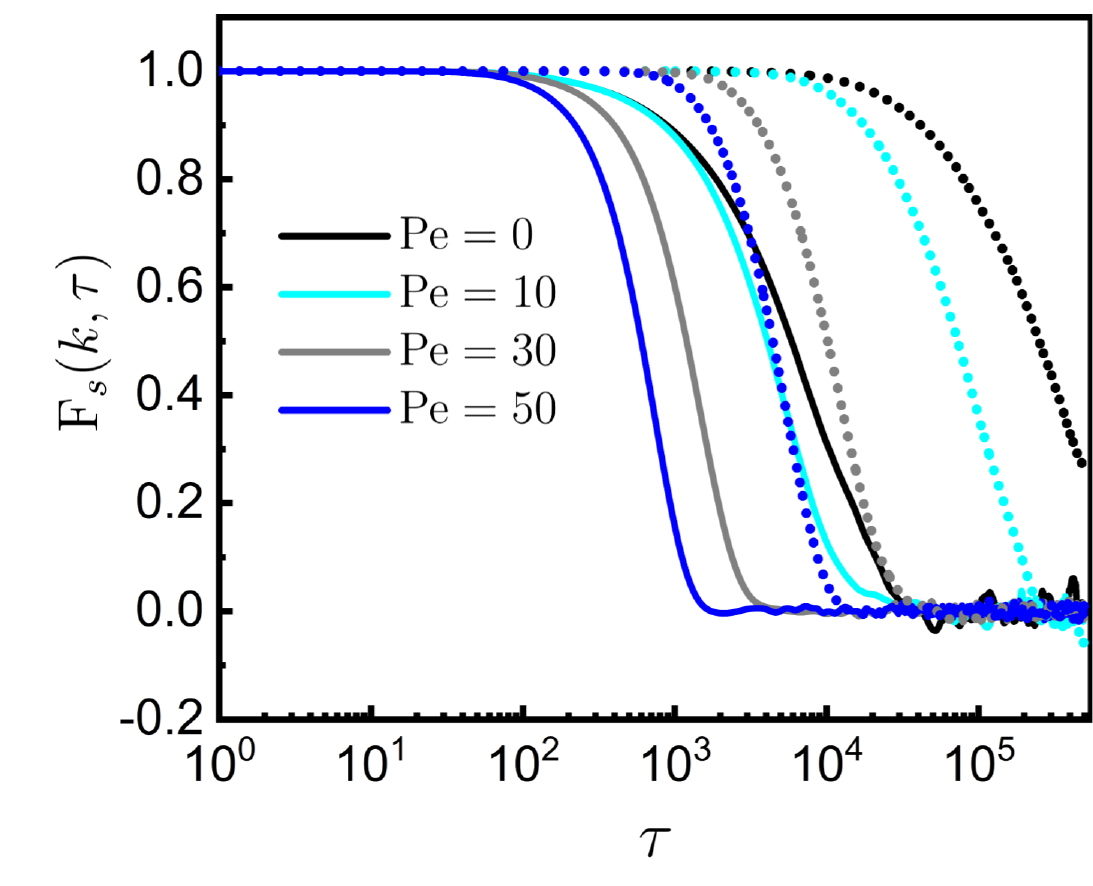}
\caption{Log-linear plot showing the time evolution of the self-intermediate
scattering function, \({\text{F}_{s}(k, \tau)}\), for different values of
\(\mathrm{Pe}\) at \({k = 1.0}\) (dotted lines) and \({k =
6.5}\) (solid lines).}
\label{fig:images3}
\end{figure}

For Brownian dynamics, the intermediate scattering function (ISF) is given
by $\mathrm{F}_{s}({k},\tau)=\exp(-k^2D\tau)$, i.e., the ISF
relaxes faster at higher $k$ values. To understand the dynamics of
the center-of-mass motion of a passive star polymer under the influence of
the active bath, we plot $\mathrm{F}_{s}({k},\tau)$ at $k=1.0$
(dotted lines) and $k=6.5$ (solid lines) in Fig.~\ref{fig:images3}
for different values of $\mathrm{Pe}$. We observe that, for a given $\mathrm{
Pe}$, the ISF relaxes faster at $k=6.5$ as compared to $k=
1.0$.

\section{Radial distribution function of active Brownian particles surrounding
linear and star polymers}
\label{appd}

\begin{figure*}
\includegraphics[width=0.70\linewidth]{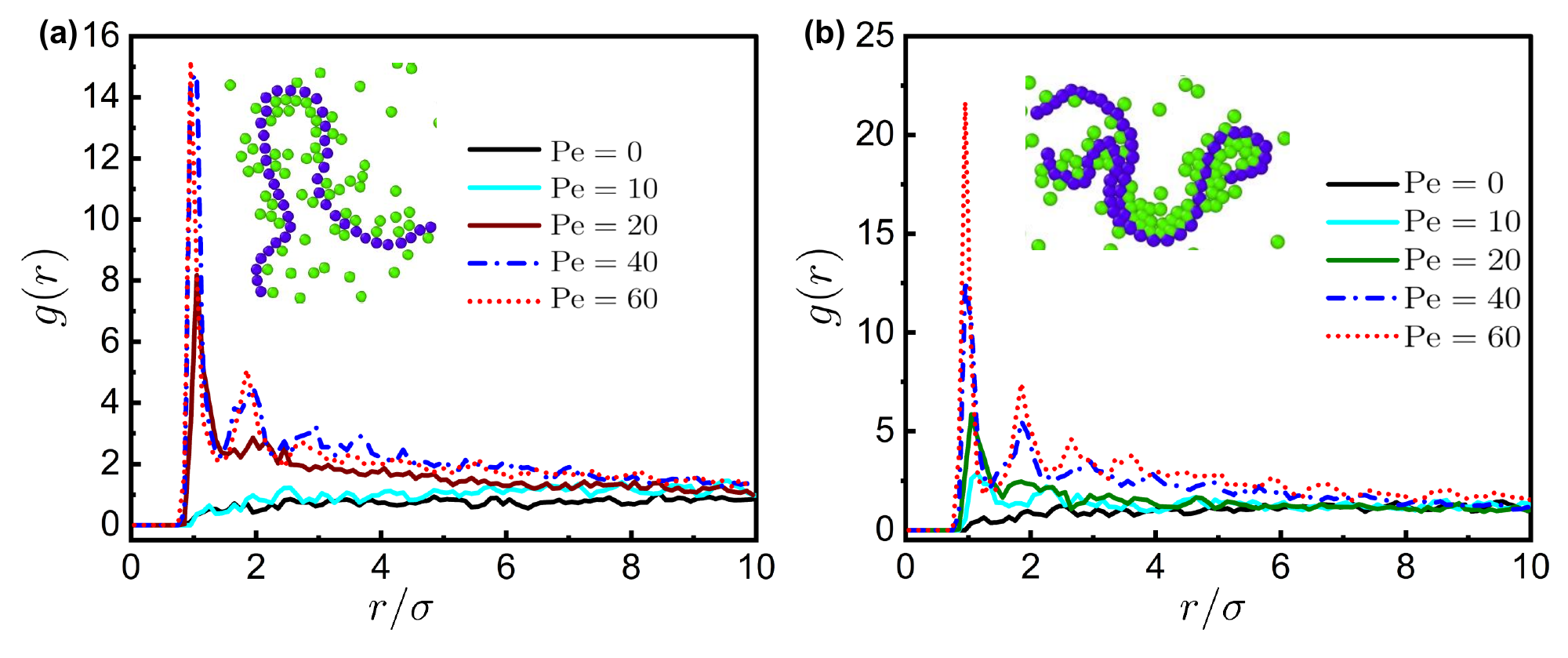}
\caption{Radial distribution function of ABPs surrounding (a) a linear polymer
and (b) a star polymer. The insets in both panels show snapshots of the
respective polymers surrounded by ABPs in an active bath with $\mathrm{Pe}=60$.}
\label{fig:images4.png}
\end{figure*}

In the case of a linear polymer, prolonged persistent motion is related to
the looping of the polymer ("parachute shape") caused by the accumulation of
active particles \cite{shin2015facilitation}. To understand the effect of the
accumulation of active particles near our star polymer and how this differs
from linear polymers, we calculated the radial distribution function
$g(r)$, of the ABPs surrounding the polymer filaments, as
shown in Fig.~\ref{fig:images4.png}. We observe that in both cases---linear
polymer in Fig.~\ref{fig:images4.png}(a) and star polymer
Fig.~\ref{fig:images4.png}(b)---there is a noticeable accumulation of ABPs
around the polymer. This accumulation becomes more pronounced at higher
activity, reflected by an increase in the peak height of $(g(r))$ at $r=1$ with increasing $\mathrm{Pe}$. Furthermore,
we observe that with increasing activity, a secondary peak emerges at larger
distances. This secondary peak is more pronounced for the star polymer as
compared to the linear polymer. From the snapshots shown as insets in the
two panels of Fig.~\ref{fig:images4.png}, we observe that ABP accumulation
occurs primarily in regions of higher curvature, which favors a similar
structure for both the linear and star polymer at high $\mathrm{Pe}$.

\section{Difference in the squared radius of gyration of linear and star
polymers as a function of activity}
\label{appe}

\begin{figure*}
\includegraphics[width=0.70\linewidth]{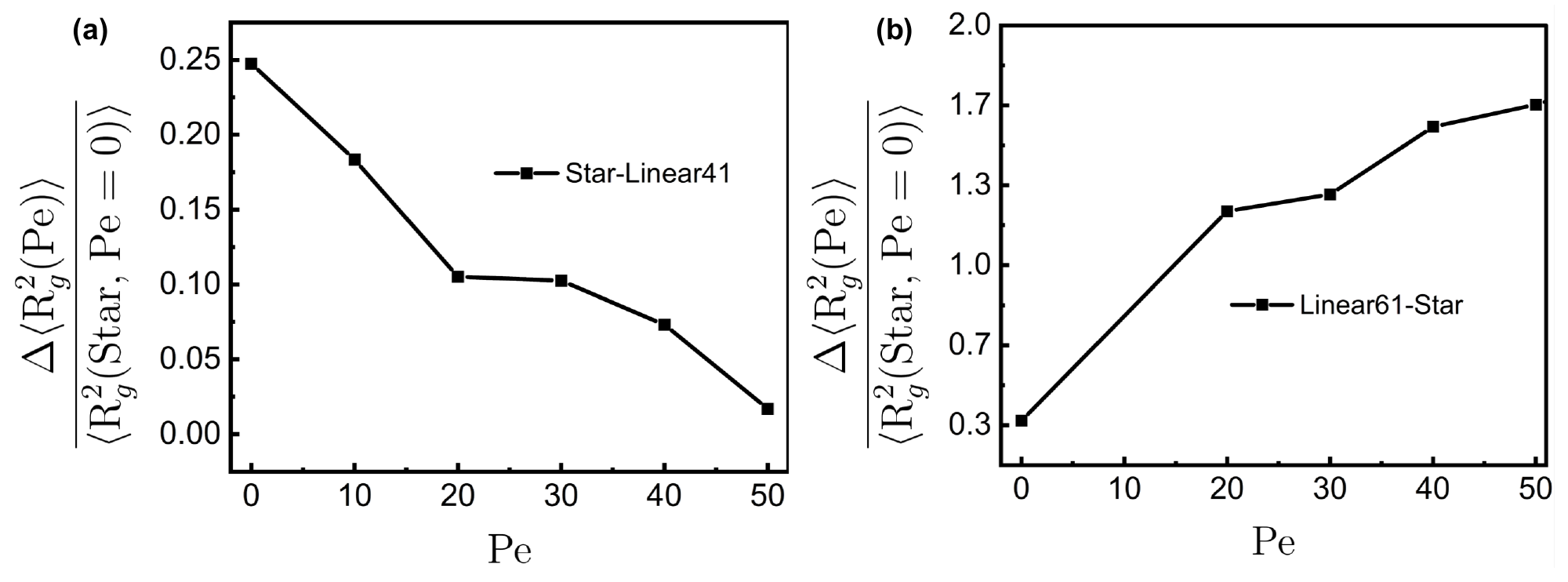}
\caption{Relative difference of the average squared radius of gyration as a
function $\mathrm{Pe}$ for (a) the star polymer and a linear polymer with 41
beads, and (b) the star polymer and a linear polymer with 61 beads.}
\label{fig:images5.png}
\end{figure*}

Here, we present a plot showing the relative difference in the average squared
radius of gyration $\frac{\Delta\langle\mathrm{R}_g^2(\mathrm{Pe})\rangle}{
\langle\mathrm{R}_g^2(\mathrm{Star,Pe}=0)\rangle}$, for a star polymer and a
linear polymer with 41 beads in Fig.~\ref{fig:images5.png}(a), and for a
linear polymer with 61 beads and a star polymer in Fig.~\ref{fig:images5.png}(b).
For the star polymer and the linear polymer with 41 beads, the relative
difference decreases with increasing $\mathrm{Pe}$, whereas in the case of
the linear polymer with 61 beads and the star polymer, it increases.

\section{Description of the movies}
\label{appf}

\textbf{Movie\_S1:} The movie demonstrates that the passive star polymer (blue)
adopts a coil conformation in a bath of ABPs (red) with $\mathrm{Pe}=0$ and
$\phi=0.06$.

\textbf{Movie\_S2:} The movie shows that the passive star polymer (blue), in
a bath of ABPs (red) with $Pe=30$ and $\phi=0.06$, adopts a swollen
conformation. Additionally, we observe an accumulation of ABPs between the
arms, facilitating their separation.

\textbf{Movie\_S3:} The movie shows that the passive star polymer (blue), in a
bath of ABPs (red) with $\mathrm{Pe}=50$ and $\phi=0.06$, adopts an interesting
conformation where two of its arms are paired while the third remains at a distance.

\bibliography{passivepol}

\end{document}